\documentclass[nofootinbib,a4paper,10pt,superscriptaddress]{openjournal}

\usepackage{amsmath,amssymb,graphicx,mathrsfs}
\usepackage{appendix,bm,dcolumn}
\usepackage[T1]{fontenc}
\usepackage[utf8]{inputenc}
\usepackage{color}
\usepackage[unicode=true,pdfusetitle,bookmarks=true,bookmarksnumbered=false,bookmarksopen=false,breaklinks=true,pdfborder={0 0 0},backref=false,colorlinks=true]{hyperref}
\hypersetup{citecolor=blue,filecolor=blue,linkcolor=blue,urlcolor=blue}
\usepackage{braket}

\newcommand{\sayy}[1]{`#1'}
\def\Eu{ \mathfrak{H} }

\def\bea{\begin{eqnarray}}
\def\eea{\end{eqnarray}}

\newcommand{\mesc}{\texttt{mescaline}}
\newcommand{\lcdm}{$\Lambda {\rm CDM}$}
\newcommand{\alp}{\alpha}

\newcommand{\ph}{\phantom}
\newcommand{\hp}{\texttt{healpy}}

\begin{document}

\title{A theoretical prediction for the dipole in nearby distances using cosmography}

\author{Hayley J. Macpherson}
\email{hjmacpherson@uchicago.edu}
\affiliation{Kavli Institute for Cosmological Physics, The University of Chicago, 5640 South Ellis Avenue, Chicago, Illinois 60637, USA}
\affiliation{NASA Einstein Fellow}

\author{Asta Heinesen}
\email{asta.heinesen@nbi.ku.dk}
\affiliation{Niels Bohr Institute, Blegdamsvej 17, DK-2100 Copenhagen, Denmark}

\begin{abstract} 
Cosmography is a widely applied method to infer kinematics of the Universe at small cosmological scales while remaining agnostic about the theory of gravity at play. Usually cosmologists invoke the Friedmann-Lemaitre-Robertson-Walker (FLRW) metric in cosmographic analyses; however, generalised approaches allow for analyses outside of any assumed geometrical model. These methods have great promise to be able to model-independently map the cosmic neighborhood where the Universe has not yet converged to isotropy. In this regime, anisotropies can bias parameter inferences if they are not accounted for, and thus must be included for precision cosmology analyses, even when the principal aim is to infer the background cosmology. In this paper, we develop a method for predicting the dipole in luminosity distances that arises due to nearby inhomogeneities. This is the leading-order correction to the standard isotropic distance-redshift law. Within a very broad class of general-relativistic universe models, we provide an interpretation of the dipole in terms of the gradients in expansion rate and density which is free from any underlying background cosmology. We use numerical relativity simulations, with improved initial data methods, alongside fully relativistic ray tracing to test the power of our prediction. We find our prediction accurately captures the dipole signature in our simulations to within $\sim 10\%$ for redshifts $z\lesssim 0.07$ in reasonably smooth simulations. In the presence of more non-linear density fields, we find this reduces to $z\lesssim 0.02$. This represents up to an order of magnitude improvement with respect to what is achieved by naive, local cosmography-based predictions. Our paper thus addresses important issues regarding convergence properties of anisotropic cosmographic series expansions that would otherwise limit their applicability to very narrow redshift ranges.

\end{abstract}
                             
\maketitle

\section{Introduction}
{In cosmology, constraints on the properties of our Universe are typically performed within a fixed model. }
The current standard $\Lambda$ cold dark matter ($\Lambda$CDM) cosmological model, for instance, adopts a general-relativistic Friedmann-Lema\^itre-Robertson-Walker (FLRW) model universe with ordinary matter, cold dark matter, radiation, and dark energy.
With these basic ingredients, cosmologists %
then use observational datasets to constrain %
{parameters describing the expansion and geometry within this model.}
While this approach has enjoyed great success in building a concordance cosmological model over many decades, $\Lambda$CDM is not free from issues. Both the unknown origin of the dark sector and the increasing prevalence of cosmological ``tensions'' \citep{Perivolaropoulos:2021jda,Abdalla:2022yfr} indicate that this model has room for improvement. %
{Such issues} have proven very difficult to solve by modifying the stress-energy sector of the $\Lambda$CDM model \citep{Schoneberg:2021qvd,Anchordoqui:2021gji}.
{Developing model-independent approaches for data analysis may help to solve some of these cosmological puzzles.}

A different route to constrain the space-time geometry is 
\textit{cosmography}: 
a purely geometrical approach to data analysis {based on series expansions of observables, such as the distance-redshift relation. This approach allows us to} 
{study} the kinematics and curvature of the Universe without {needing to assume} 
anything about the underlying theory of gravity 
\citep{Weinberg:1972kfs,Visser:2004bf}. 
Cosmography tools {are} %
widely used for analyzing supernovae of {Type 1a}, %
{in particular} in modern analyses for constraining the Hubble constant with %
low-redshift data 
\citep{Riess:2016,Riess:2019cxk,CSP:2018rag}, and {have also been} used to analyze {gravitational wave} standard sirens \citep{deSouza:2021xtg} {and} 
redshift drift \citep{Lobo:2020hcz}. 
Conventional cosmographic analyses  
{like these typically enforce} the FLRW geometry {in the distance-redshift relation}. 
{However,} cosmography is in practice applied in the regime of low redshifts \citep{Cattoen:2007sk}, where \textit{sufficient} convergence to homogeneity and isotropy is not guaranteed. 
On these scales close to the homogeneity scale of the Universe, we lose potentially valuable information by restricting the metric to be FLRW \citep{Macpherson:2021gbh,Kalbouneh:2024szq,Kalbouneh:2024yjj} and can even get biased constraints on parameters by neglecting inhomogeneities which impact our observations \citep{Kalbouneh:2022tfw,Macpherson:2024zwu,Kalbouneh:2024yjj}.

In conventional cosmography approaches, cosmic structures are either neglected (assuming that {fluctuations due to inhomogeneities} will be subdominant relative to the error bars of the final results) or accounted for by applying model-dependent corrections---such as peculiar velocity corrections \citep{Peterson:2022}---with the aim  
to \sayy{map} the data into the  
fictitious homogeneous and isotropic space-time. 
However, as 
cosmological 
data approaches sub-percent precision,   
inhomogeneities will play an increasingly important role in our measurements.  
In particular, among the tensions that are increasing in significance in modern cosmology 
lie several which are directly related to anisotropies in data and potentially larger-than-expected cosmic structures {\citep[see, e.g.][]{Aluri:2023}.}

Anisotropic cosmography provides a generalization of the FLRW cosmography framework, where the impact of cosmic structures on observables can be accounted for in a model-independent and geometrically consistent way \citep{1985PhR...124..315E}. 
Such approaches are agnostic to both the field theory of gravity \emph{and} the geometry of the Universe{; without any} 
assumptions about space-time symmetries. 
Early theoretical contributions to the field of anisotropic cosmography in general space-time geometries were made by \citet{Ehlers1961_GRG1993,Kristian:1965sz,1970CMaPh..19...31M,Ellis1971_2009GRG,1985PhR...124..315E}, and these were later developed into full comprehensive frameworks for doing cosmography  \citep{Clarkson:2010,Clarkson:2011uk,Clarkson:2011br,Umeh:2013UCT} with recent theoretical developments in, e.g., \citep{Heinesen:2020bej,Heinesen:2021qnl,Maartens:2023tib,Heinesen:2024npe,Kalbouneh:2024yjj,Sarma:2025yfw}, as well as advances in prediction studies with simulations \cite{Heinesen:2021azp,Macpherson:RT,Adamek:2024hme} and proof-of-principle analyses using {supernova} 
and galaxy data \citep{Dhawan:2022lze,Cowell:2022ehf,Kalbouneh:2025jnp}.

These generalised frameworks may help in particular in addressing the current `dipole tension': several independent observational probes have been shown to contain significantly larger dipole anisotropies than expected based on $\Lambda$CDM predictions from the cosmic microwave background (CMB). The CMB contains a large dipole anisotropy which is usually interpreted as solely due to our motion with respect to the large-scale `cosmic rest frame' \citep{Planck:2014dip}. This motion should also imprint a dipole %
{modulation} in other, lower-redshift sources. 
However, independent studies probing {dipoles in the distance-redshift relation} and galaxy number counts at various scales have
{extracted a velocity from the dipole} %
which is discrepant with the CMB at high significance \citep[e.g.][]{Migkas:2020fza,Secrest:2021,Horstmann:2022,Secrest:2022,Singal:2022,Watkins:2023,Dam:2023,Wagenveld:2023} 
though others find agreement with $\Lambda$CDM \citep[e.g.][]{Darling:2022,Dhawan:2022lze,Sorrenti:2023,Ferreira:2024,Abghari:2024,Tiwari:2024,Wagenveld:2025}. 
From the generalised cosmography, we naturally expect a dominant dipolar anisotropy in low-redshift distances due to structures which is {separate}{---}and in principle distinguishable from{---}the dipole induced by {our} peculiar motion 
\citep[see Section~5 of][]{Heinesen:2021azp}.

In this paper, we set out to define a robust prediction for the dipole in distance--redshift data based on cosmography as truncated at third order in redshift. 
We adopt a realistic universe model that is independent of any FLRW background space-time, which in general will allow for interpretation of the measured dipole in terms of gradients in expansion rate and matter density of the inhomogeneous space-time. We test our prediction using fully relativistic cosmological simulations with ray tracing which also does not explicitly enforce an FLRW background cosmology.

In Section~\ref{sec:cg}, we review the general cosmography method and make a prediction for the dipole in the luminosity distance--redshift cosmography based on this framework using an approximation scheme suitable for the late Universe. 
In Section~\ref{sec:sims}, we review the simulations and the initial data {we use} to create a realistic model universe 
to test our prediction for the {dipole and} in Section~\ref{sec:rt} we review the ray-tracing methods {we use} to calculate exact distances and redshifts within our simulations. 
In Section~\ref{sec:results}, we {discuss} our results {and we conclude} in Section~\ref{sec:conclusion}. 
{Unless otherwise stated, Greek letters are space-time indices and take values $0\ldots3$ and Latin letters are spatial indices with values $1\ldots 3$, with repeated indices implying summation, and we set the speed of light $c=1$ throughout.}

\section{Cosmography}\label{sec:cosmography} 
In Section~\ref{sec:cg} we review the cosmography for the luminosity distance--redshift relation for a general space-time and congruence description of emitters and observers.  
In Section~\ref{sec:dipcg} we describe our framework to predict the dipolar signature in the luminosity distance by smoothing the cosmography under the ``quiet universe'' approximation. 

\subsection{Review of general luminosity distance cosmography}\label{sec:cg}

The aim of this section is to arrive at a {cosmographic} expression for the luminosity distance as a function of redshift within {a generic} space-time and {for} any observer and emitters within that space-time. {This expression will form the basis of our theoretical dipole prediction. }
To this end, we consider a general 
4--velocity field $u^\mu$ as representing the time-direction of the emitters and observers in the space-time with a general space-time metric $g_{\mu\nu}$. 
Each observer measures incoming photons that move on geodesics {and are} emitted from sources that are also moving with $u^\mu$. {We assume the} family of observers and emitters %
form a smooth \textit{congruence} throughout the space-time. The redshift of the photons from each source is given by $1+z\equiv E_s/E_o$, where $s$ denotes the event of emission of the photons by the source and $o$ denotes the event of observation. The energy function of the photons is $E\equiv - u_\mu k^\mu$ where $k^\mu \equiv d x^\mu/d\lambda$ is the 4--momentum of the photons and $\lambda$ is the affine parameter of the geodesic.

We can generally approximate the luminosity distance using a Taylor series expansion in redshift within the radius of convergence of the series\footnote{For expanding FLRW models, the radius of convergence is often as large as $z=1$ \cite{Cattoen:2007sk}, but these results do not necessarily extend to universe models that are not maximally spatially symmetric, and we must in principle test the convergence for the given space-time at hand.}. 
In the formalism we adopt here, this is done by use of the Sachs optical equations and by assumption of Etherington's reciprocity theorem. 
The resulting generalized cosmographic expansion of the luminosity distance, $d_L$, in redshift, $z$, around the point of observation is \citep[see][for details]{Heinesen:2020bej} 
\begin{equation}\label{eq:series}
    d_L(z) =  d_L^{(1)} z   + d_L^{(2)} z^2 +  d_L^{(3)} z^3 + \mathcal{O}( z^4),
\end{equation} 
where $z$ lies within the radius of convergence.
The coefficients are 
\begin{equation}
    \begin{aligned} \label{eq:dLexpand2}
    d_L^{(1)} &= \frac{1}{\Eu_o} \, , \qquad d_L^{(2)} =   \frac{1 - \mathfrak{Q}_o }{2 \Eu_o}  \ ,\\% \qquad 
    d_L^{(3)} &=  \frac{- 1 +  3 \mathfrak{Q}_o^2 + \mathfrak{Q}_o    -  \mathfrak{J}_o   + \mathfrak{R}_o }{ 6  \Eu_o}     \, , 
    \end{aligned}
\end{equation} 
where a subscript $o$ means that quantity is evaluated at the point of observation. The effective Hubble, deceleration, curvature, and jerk parameters appearing in the coefficients above are defined as, respectively,
\begin{subequations}\label{eq:paramseff}
    \begin{align}
    \label{eq:hdef}
    \Eu &\equiv - \frac{1}{E^2}     \frac{ {\rm d} E }{{\rm d} \lambda}  \, ,\\ %
    \mathfrak{Q}  &\equiv - 1 - \frac{1}{E} \frac{     \frac{ {\rm d} \Eu}{{\rm d} \lambda}    }{\Eu^2}   \, , \\ %
    \mathfrak{R} &\equiv  1 +  \mathfrak{Q}  - \frac{1}{2 E^2} \frac{k^{\mu}k^\nu R_{\mu \nu} }{\Eu^2}   \, , \\% \qquad
    \mathfrak{J}  &\equiv   \frac{1}{E^2} \frac{      \frac{  {\rm d^2} \Eu}{{\rm d} \lambda^2}    }{\Eu^3}  - 4  \mathfrak{Q}  - 3 \, ,
    \end{align}
\end{subequations}
where $R_{\mu\nu}$ is the Ricci tensor of the space-time. These parameters can be thought of as generalizations of the FLRW parameters of the same name; though now depending 
on both the \textit{position} of the observer and the \textit{direction} of observation. 
These generalized cosmological parameters contain information about the structures and curvature in the space-time, and as opposed to the conventional cosmological parameters defined for the FLRW geometries only, they are defined for \emph{any} space-time model with a congruence description for the sources within it. 
For instance, in the special case of a linearly-perturbed FLRW model, they will contain information about the density contrast and the peculiar velocity field of the matter, and in the special case of a Bianchi I cosmology they will contain information about the shear tensor describing the anisotropic deformation of space in the frame of the co-moving observers. 
The generalized cosmological parameters are formulated in terms of expansion and curvature degrees of freedom of the underlying space-time, as detailed in e.g. \citet{Heinesen:2020bej,Kalbouneh:2024yjj}, and one can thus learn about the expansion and curvature of the Universe directly from fitting the above cosmography to data.
In deriving the expressions above, there have been no assumptions on the form of the metric tensor of space-time or the field equations governing the evolution of that space-time, and the cosmography expressions may thus be applied to any space-time as long as geometrical optics applies and that the luminosity distance as a function of redshift is an analytical function. 

Importantly, the generalised parameters \eqref{eq:paramseff} can be written as multipole series with a finite number of coefficients. 
This implies that, so long as the third-order truncation {in redshift} of %
\eqref{eq:series} remains accurate for the {data,}
this formalism can be used {to constrain curvature and expansion degrees of freedom, even in scenarios that go significantly beyond the FLRW model.}
We refer the reader to \citet{Heinesen:2020bej} for more details on the derivation of this formalism and multipole decompositions of the parameters \eqref{eq:paramseff}. 
See also important earlier work on covariant cosmography by \citet{Kristian:1965sz,1970CMaPh..19...31M,1985PhR...124..315E,Clarkson:2011uk,Clarkson:2011br,Umeh:2013UCT} leading up to this result.

This generalised cosmographic framework forms the central component in our new method to predict the dipole in distances which arises due to inhomogeneities in space-time. In the next section, we will identify the dominant contributions to the dipole component of \eqref{eq:series} by making some reasonable approximations for the space-time we are interested in.

\subsection{Dipole predicted from smoothed cosmography}\label{sec:dipcg} 
 
We aim to develop a robust prediction for the dipole which is applicable over a range of scales. 
One of the main challenges in applying cosmography in practice is that the convergence and level of approximation of a given order of the series expansion is not known when the space-time metric is left unspecified. Generally, the smaller the structures that are accounted for, the more terms in the series expansion must be included to fit the cosmic observables at a given scale \citep{Macpherson:2021gbh}. The problem is that we will in practice only be able to fit a cosmography truncated at low orders, as the number of free variables to be fitted grow rapidly with the order of the cosmography \citep{Kalbouneh:2024szq}. 
Studies of the accuracy of the third order distance-redshift cosmography 
in specific anisotropic model scenarios \citep{Macpherson:RT,Koksbang:2024nih,Modan:2024txm} confirm that the presence of high density contrasts result in {divergences and} large remainder terms even at relatively modest redshifts in the cosmological context. 

To mitigate these issues and attempt to accurately predict the dipole in distances for a broader redshift range, we implement a smoothing operation for the cosmography coefficients, such that they represent \textit{mean} kinematic and curvature degrees of freedom in the space-time at a given scale of observation, rather than  \textit{local} degrees of freedom.

For this purpose, we shall assume that we are on sufficiently large scales so that any pressure and vorticity of the matter can be ignored as well as the divergence of the magnetic Weyl tensor. For the former assumption, this condition should be satisfied above the galaxy cluster scale where dark matter can be accurately described as a vorticity-free perfect dust fluid. For the latter, this ensures that the shear and electric Weyl tensor share a common eigenbasis.

Based on the above, we thus model the dark matter as
a pressureless `dust' fluid, such that the energy-momentum tensor at leading order is $T_{\mu \nu} = \rho\,u_\mu u_\nu$, where $u^\mu$ is the 4--velocity of the fluid and  $\rho$ is the rest-mass density in the matter frame. Finally, we assume that Einstein's theory of general relativity holds\footnote{Making an assumption about the form of the field equations is strictly speaking not in the spirit of cosmography. We, however, make this leading-order assumption for the field equations in order to leave the geometry side generic.}.  
This set of assumptions corresponds to the ``quiet universe'' approximation as described and investigated in \citet{Heinesen:2021azp}. 
{The quiet universe models can be seen as generalizations of the silent universe models \citep{Bruni:1994nf,vanElst:1996zs}, where the assumption about vanishing magnetic Weyl tensor is relaxed, %
{thus allowing gravitational radiation to propagate} \citep{Matarrese:1993zf,Dunsby:1997fyr}. 
{This allows} the \sayy{sound} of gravitational waves while still being silent in terms of pressure waves. 
Intuitively, it %
{might} seem reasonable to neglect gravitational at leading order. However, 
the silent universe approximation has been shown to have a linearisation instability that makes it inappropriate for describing non-linear structures \citep{vanElst:1996zs}. In contrast, the quiet universe approximation was shown to accurately describe data from  realistic universe simulations \citep{Heinesen:2021azp} similar to those investigated in this paper, and we thus use the quiet universe approximation here.   }

The kinematic decomposition of the matter frame for the vorticity-free fluid is 
\begin{align} \label{def:expu}
        \nabla_{\nu}u_\mu  &= \frac{1}{3}\theta\, h_{\mu \nu }+\sigma_{\mu \nu} \, , \qquad \theta \equiv \nabla_{\mu}u^{\mu} \, , \qquad \sigma_{\mu \nu} \equiv h^\beta_{\, ( \mu} h^\alpha_{\, \nu ) } \nabla_{\alpha} u_{\beta} - \frac{1}{3} \theta\, h_{\mu \nu} \, , 
\end{align} 
where $\theta$ is the volume expansion rate and $\sigma_{\mu \nu}$ is the volume shear rate describing the anisotropic deformation {of the matter frame}. 
In the above, $\nabla_\mu$ is the covariant derivative associated with the space-time metric $g_{\mu\nu}$, $h_{\mu\nu}=g_{\mu\nu} + u_\mu u_\nu$ is the projection tensor to the matter frame, and parentheses implies symmetrization over those indices. 
We also introduce the spatial covariant derivative, defined through its acting on a tensor field, $T^{\mu \nu ...}{}_{\sigma \rho ...}$, as
\begin{equation}
\label{eq:Diff}
D_{\kappa}T^{\mu \nu ...}{}_{\sigma \rho ...} \equiv h^\mu{}_\gamma \,h^\nu{}_\epsilon \,h^\psi {}_\sigma\,h^\chi {}_\rho\,
 ... h^\epsilon{}_\kappa \nabla_\epsilon\,T^{\gamma \epsilon ... }{}_{\psi \chi ... } \ , 
\end{equation} 
where, %
{for example,} the spatial gradient of a scalar $S$, thus reads $D_\mu S = h^\nu_{\, \mu} \nabla_\nu S$. 

We shall now make the further assumption that $\sigma^{\mu \nu} \sigma_{\mu \nu} \ll \theta^2$ and that $E_{\mu \nu} E^{\mu \nu} \ll \theta^4$, where $E_{\mu \nu}$ is the electric part of the Weyl tensor. 
However, we shall allow for the spatial divergences of $\sigma^{\mu \nu}$ and $E_{\mu \nu}$ to be of similar size as $D_\mu \theta$ and $\theta D_\mu \theta$, respectively. 
Under the quiet universe approximation, with these additional smallness assumptions, 
we can derive a prediction for the dominant terms in the coefficients of the series \eqref{eq:series}. 
We shall write the dominant dipole in the cosmographic expansion as 
\begin{equation}\label{eq:dLdip_predic}
    d_{L,{\rm dip}}({\bf e}, z) = e^\mu \left( d_{L,\mu}^{(2)} z^2 + d_{L,\mu}^{(3)} z^3 \right) , 
\end{equation} 
where $d_{L,\mu}^{(2)}$ and $d_{L,\mu}^{(3)}$ are the dipolar terms of the cosmography coefficients $d_{L}^{(2)}$ and $d_{L}^{(3)}$ in \eqref{eq:dLexpand2}, and $e^\mu$ is the direction of observation. 
The first order term $d_L^{(1)}$ contains no dipole term since the observers are co-moving with the matter frame and thus have zero acceleration \citep[see][]{Macpherson:2021gbh}. We have also neglected any small-scale induced peculiar velocity relative to the large scale matter flow \citep[e.g. motions within our galaxy or local group; see Section~5 of][for ways to account for such effects in a consistent cosmographic analysis]{Heinesen:2020bej}. 
{Although, we do} consider the prediction from a pure local motion effect {and its potential to improve the prediction at large scales.}
The dominant dipolar components of the second and third order coefficients follow from the approximations above together with the covariant  constraint and evolution equations in general-relativistic space-times \citep{1996PhDT........25V}, and are given by \citep[see][for details]{Heinesen:2021azp}

\newcommand{\bse}{\boldsymbol{e}}
\begin{subequations}\label{eqs:dLidip}
\begin{align}
    d_{L,\mu}^{(2)} &= %
    - \frac{3}{5} \frac{ D_\mu \theta}{2  (\frac{1}{3}\theta)^3}  \, , \label{eq:dL2dip} \\
    d_{L,\mu}^{(3)} &= \frac{ \frac{3}{5} \left(  \frac{16}{27} + \frac{9 \times 8 \pi G \rho}{\theta^2}\right)   D_\mu \theta - \frac{23}{15} \times 8\pi G D_\mu \rho}{6\,  (\frac{1}{3}\theta)^3 }   \, ,  \label{eq:dL3dip}
\end{align}
\end{subequations} 
where we have expanded to zeroth order in $\sigma_{\mu\nu}/\theta$ and $E_{\mu\nu}/\theta^2$ consistent with our approximation outlined above.

The source of the dipolar signal in the coefficients \eqref{eqs:dLidip} are gradients of the scalars $\theta$ and $\rho$. In a universe with highly nonlinear structures, these gradients can be large even if {fluctuations in} the metric itself remain small. Highly inhomogeneous terms such as $\theta$ and $\rho$ contribute to the difficulty in capturing distances across wide redshift ranges with the general cosmography as truncated at third order \citep{Macpherson:2021gbh}. 
With a view towards a robust cosmographic prediction applicable across a \textit{range} of redshifts, we shall consider the expression for the dipole in terms of \textit{smoothed} values of $\theta$ and $\rho$. 
We thus consider the expressions in \eqref{eqs:dLidip} with $\theta \mapsto \braket{\theta}$ and $\rho \mapsto \braket{\rho}$, i.e. %
\begin{subequations}\label{eqs:dLidip_av}
\begin{align}
    d_{L,\mu}^{(2)} &= %
    - \frac{3}{5} \frac{ D_\mu \braket{\theta} }{2  (\frac{1}{3}  \braket{\theta})^3}  \, , \label{eq:dL2dip_av} \\
    d_{L,\mu}^{(3)} &= \frac{ \frac{3}{5} \left(  \frac{16}{27} + \frac{9 \times 8 \pi G \braket{\rho}}{\braket{\theta}^2}\right)   D_\mu \braket{\theta} - \frac{23}{15} \times 8\pi G D_\mu \braket{\rho}}{6\,  (\frac{1}{3}\braket{\theta})^3 }   \, ,  \label{eq:dL3dip_av}
\end{align}
\end{subequations} 
where $\braket{}$ represents a volume average in some region surrounding the observer. 
For non-linear fields, we will in general have substantial variance such that  $\braket{\theta^2} \neq \braket{\theta}^2$ which will make the mapping from \eqref{eqs:dLidip} to \eqref{eqs:dLidip_av} that we have made ambiguous. The fact that averages and non-linear operations do not commute could bias predictions made based on the mapping suggested both in models with and without backreaction. For the purposes of this initial investigation, we will assume that the mapping $\theta \mapsto \braket{\theta}$ and $\rho \mapsto \braket{\rho}$ holds for the purpose of predicting observables at lowest order for the model universe of interest.

{We perform the smoothing} over a spatial scale corresponding roughly to the redshift of observation. 
We smooth $\theta$ and $\rho$ using a Gaussian kernel assuming spatial flatness, 
with specifics described in Section~\ref{sec:dipoleprediction}.
{An exception is for spatial scales which correspond to the size of one grid cell for the simulations: at this scale we expect the local cosmography description \eqref{eqs:dLidip} to be very good and the addition of smoothing in fact makes the prediction worse.}

In principle, the smoothing could instead be performed using a more accurate geometric averaging procedure such as is described in \citet{Buchert:2000,Buchert:2001sa,Buchert:2020a}. However, the Gaussian smoothing procedure outlined in Section~\ref{sec:dipoleprediction} %
is sufficient for the simulations employed in this work {given that scalar fluctuations in the metric remain small \citep{Macpherson:2019a}}.

\section{Simulations} \label{sec:sims}

In Section~\ref{sec:ET}, we describe the 
software and initial data 
for the model universe that we use as a test case in this analysis. 
In Section~\ref{sec:dipoleprediction}, we describe our computation of the dipole prediction \eqref{eqs:dLidip_av} based on the simulation output. 

\subsection{Numerical relativity software framework} \label{sec:ET}

We use the Einstein Toolkit\footnote{\url{https://einsteintoolkit.org}} \citep[ET;][]{EinsteinToolkit:2019_10,Loffler:2012}, an open-source numerical relativity (NR) code based on the Cactus\footnote{\url{https://www.cactuscode.org}} infrastructure. NR is a computational method that allows us to evolve Einstein's field equations free from space-time symmetries given a set of consistent initial data. We use the Baumgarte-Shapiro-Shibata-Nakamura \citep[BSSN;][]{Baumgarte:1999,Shibata:1995} formulation of NR which makes use of a 3+1 decomposition of space-time with adapted space-time coordinates $x^\mu = (t,x^i)$. 
Spatial surfaces are defined by their normal vector $n^\mu$ which is decomposed into the lapse function $\alpha\equiv - n^\mu \partial_\mu t$---describing the spacing in time between spatial slices---and the shift vector $\beta^i\equiv n^\mu \partial_\mu x^i$---describing the shift in spatial coordinates between the slices. These functions parameterise the coordinate degrees of freedom of GR and are thus arbitrary choices which should not impact observable results from the simulations. The specific gauge choices we make in this work are described in detail in \citet{Macpherson:2019a}.

Our cosmological initial data is generated using \texttt{FLRWSolver} \citep[see Section~\ref{sec:ICs} of][]{Macpherson:2017}, the space-time is evolved using the \texttt{McLachlan} thorn \texttt{ML\_BSSN} \citep{Brown:2009}, and the hydrodynamics is evolved using \texttt{GRHydro} \citep{Baiotti:2005}. Our simulations are matter dominated, i.e. contain no dark energy, which is a choice purely for convenience given that there is currently no option to include a cosmological constant in the ET. 
The predictions made in Section~\ref{sec:dipcg} hold when a cosmological constant is included, and we thus expect the main results of our analysis assessing this prediction to hold in the presence of a cosmological constant. 
The matter content is described as a pressureless `dust' fluid, which we implement in \texttt{GRHydro} by ensuring $P\ll \rho$ via our choice of equation of state in \texttt{EOS\_Omni}. This setup has been successful in accurately reproducing exact dust solutions within perturbed FLRW solutions %
\citep{Macpherson:2017}, and has been used to study nonlinear matter distributions \citep[e.g.][]{Macpherson:2019a}.

\subsection{Initial data}\label{sec:ICs}

Initial data as set with \texttt{FLRWsolver} are perturbations to a background Einstein-de Sitter (EdS) space-time determined by a user-supplied initial power spectrum at the chosen initial redshift. 
Previous works using \texttt{FLRWSolver} used the linear perturbation option: the initial density field specified by an input power spectrum and the resulting velocity and metric perturbations determined via linear perturbation theory \citep[e.g.][]{Macpherson:2019a,Macpherson:2021gbh,Macpherson:2024zwu}. Here we use the new `exact' perturbation option: the initial \textit{metric} perturbation field specified by an input power spectrum and the resulting density and velocity perturbations determined `exactly' by solving the Hamiltonian and momentum constraints. 
In the 3+1 decomposition adopted in BSSN, Einstein's equations are cast into a set of evolution equations for the metric describing the spatial hypersurfaces and a set of \textit{constraint equations}. These equations contain no time dependence and so must be satisfied on every spatial slice for any solution of Einstein's equations. Violations to these constraints are defined as
\begin{subequations}\label{eqs:constraints}
\begin{align}
    H &\equiv \mathcal{R} - K_{ij}K^{ij} + K^2 - 16\pi G \rho,\label{eq:Hconstraint} \\
    M_i &\equiv D_j K^j_{\phantom{j}i} - D_i K - 8\pi G S_i,\label{eq:Mconstraint}
\end{align}
\end{subequations}
where $\mathcal{R}\equiv \gamma^{ij}\mathcal{R}_{ij}$ with $\gamma_{ij}$ the spatial metric tensor and $\mathcal{R}_{ij}$ the 3--Ricci tensor, $K_{ij}$ is the extrinsic curvature of the spatial surfaces (with $K$ its trace), $D_i$ is the covariant derivative associated with the spatial metric as defined in \eqref{eq:Diff},   
and $\rho\equiv T_{\mu\nu}n^\mu n^\nu$ and $S_i=\gamma_{i\alp}n_\beta T^{\alp\beta}$ are the energy and momentum densities of the hypersurfaces, respectively. 
{We will have $H=0$ and $M_i=0$} for exact solutions {of Einstein's equations}, however for numerical simulations {these} are in general non-zero due to the finite accuracy of numerical schemes. Typically \eqref{eqs:constraints} are used to set initial data, sometimes under additional simplifying assumptions that also contribute to a violation of this system (as for linear perturbations with \texttt{FLRWSolver}, where the violation is second order in the perturbation amplitudes).

In the new `exact' option for initial data in \texttt{FLRWSolver} we define an initial parameterisation of the metric as
\begin{equation}\label{eq:pertmetric}
    ds^2 = - a(t)^2 \,(1 + 2 \,\phi)\, dt^2 + a(t)^2\,(1-2\phi)\,\delta_{ij}\, dx^i \, dx^j,
\end{equation}
where $a(t)$ is the background scale factor, $\phi(x^i)$ is the Newtonian potential, and $\delta_{ij}$ is the Kronecker delta function. 
\texttt{FLRWSolver} generates an initial $\phi(x^i)$ in Fourier space as Gaussian-random fluctuations following a user-input power spectrum $P_\phi(k)$. 
The metric \eqref{eq:pertmetric} is then defined with this $\phi$ and a scale factor defined from the chosen initial background redshift as $a_{\rm ini}=1/(1+z_{\rm ini})$.

From this metric, \texttt{FLRWSolver} calculates the Ricci tensor and extrinsic curvature, plus its derivatives, which constitute the space-time terms of \eqref{eqs:constraints}. 
For $K_{ij}$, we further assume $\partial\phi/\partial t \equiv \partial_t \phi = 0$ which is a good approximation when both the amplitude of $\phi$ is small and during the matter-dominated era of the universe. We can then invert \eqref{eqs:constraints} to simply obtain the densities $\rho$ and $S_i$. 
\texttt{FLRWSolver} communicates with the ET thorn \texttt{HydroBase} which stores the \textit{rest-mass density} $\rho_0\equiv T_{\mu\nu}u^\mu u^\nu$ and the fluid velocity with respect to the Eulerian observer $v^i \equiv u^i / (\alp u^0)$. We thus need to extract these quantities from the hypersurface-frame densities $\rho$ and $S_i$.
We can write $\rho$ in terms of the rest-mass density as $\rho=\rho_0\, \alp^2 (u^0)^2=\rho_0 \Gamma^2$ where $\Gamma=1/\sqrt{1-v^i v_i}$ is the Lorentz factor describing the tilt between the fluid frame and the hypersurfaces, and the momentum density as $S^i = \alp \rho_0 u^i u^0$. Thus, we can obtain the rest-mass density and velocity field via \citep[as in][]{Adamek:2020}
\begin{equation}   
   \rho_0 = \rho - \frac{S^i S_i}{\rho}; \quad\quad v^i = \frac{S^i}{\rho}.
\end{equation}

With this method, we have not made any perturbative approximations for solving Einstein's equations, 
and thus the constraint violation on the initial slice is determined by the fourth-order accuracy of the finite difference derivative approximation we use to calculate the Ricci tensor from \eqref{eq:pertmetric}.

The initial power spectrum 
$P_\phi(k)$ {can be} output from, e.g., CAMB\footnote{\url{https://camb.info}} \citep{Lewis:2002} or CLASS\footnote{\url{https://lesgourg.github.io/class_public/class.html}} \citep{Lesgourgues2011,CLASS:2011} {and supplied to \texttt{FLRWSolver}. It is important to note, however,} that these codes assume a linearised FLRW evolution to propagate {the perturbations} to the chosen redshift of the {desired} power spectrum. 
\texttt{FLRWSolver} {in principle} acts independently of {the origin of the supplied initial} power spectrum, and the resulting initial data will be accurate to all orders of that metric perturbation. The main benefit of this method is that it allows us to initialise the simulations at lower redshifts, i.e. down to $z_{\rm ini}=20$ instead of $z_{\rm ini}=1000$ as in previous works \citep[e.g.][]{Macpherson:2019a,Macpherson:2021gbh,Macpherson:2024zwu}. This reduces the number of timesteps in the simulation and thus reduces the accumulation of finite-difference error in the constraint violation. {This yields} about a factor of two improvement {in constraint violation at $z=0$} with respect to earlier {simulations}.  
{This improvement is currently available in the ET as a part of the May 2025 ``Martin D. Kruskal'' release.}

\subsection{Simulations used in this work}

In this work, we analyse a set of simulations initialised {with an EdS background} 
redshift of $z_{\rm ini}=20$ and evolved until the 
volume of the domain has increased by a factor of {$(1+z_{\rm ini})^3$}. 
These simulations have co-moving length $L=2.56\,h^{-1}$ Gpc and $L=3.072\,h^{-1}$ Gpc.
The initial power spectrum $P_\phi(k)$ is output from CAMB using $h=0.45$ (where $h\equiv H_0/100$ km/s/Mpc is the dimensionless Hubble parameter) and otherwise default Planck parameters \citep{PlanckCosmo:2020}. We use three simulations with $L=2.56\,h^{-1}$ Gpc and resolutions $N=64, 128$, and 256---all with identical initial data---{which we use} to assess the numerical convergence of our results {(see Appendix~\ref{appx:conv})}, and one simulation with $L=3.072\,h^{-1}$ Gpc and resolution $N=256$. {We assess the level of violation in the Hamiltonian constraint for all simulations in Appendix~\ref{appx:constraint}.}

Our NR simulations are performed with the matter content discretised on a cubic grid rather than particles{, the latter of} which is more common in Newtonian cosmological simulations. 
This limits our ability to simulate structures below the scale of bound objects, i.e. galaxy clusters, given the breakdown of the continuous fluid approximation at this scale. We thus ensure our simulations never sample scales below $\sim 8\,h^{-1}$ Mpc by ensuring our grid cells are always $\gtrsim 4\,h^{-1}$ Mpc in size {(since the Nyquist frequency is the scale corresponding to two grid cells)}. Additionally, our discretised grid means that the smallest-scale modes are not {necessarily} sampled with sufficiently many points, which introduces significant numerical error especially when these modes become nonlinear. To mitigate this, we remove small-scale modes from the initial data to delay the onset of this additional numerical error. Specifically, 
we perform a cut to the initial power spectrum %
such that $P(k>k_{\rm max})=0$, where $k_{\rm max}=2\pi/\lambda_{\rm min}$ and $\lambda_{\rm min}$ %
is the wavelength of the smallest mode sampled in the simulation. 
For equivalent initial data across three resolutions, this means the power spectrum cut must be quite large to ensure each simulation has its minimum mode sampled by {at least} 10 grid cells. The minimum scale sampled in the simulations with $L=2.56\,h^{-1}$ Gpc is thus $\lambda_{\rm min}=400\,h^{-1}$ Mpc {(i.e., 10 grid cells for the lowest resolution $N=64$)}. For the simulation with $L=3.072\,h^{-1}$ Gpc, we maintain a minimum scale cut of 10 grid cells for $N=256$, yielding $\lambda_{\rm min}=120\,h^{-1}$ Mpc, which means this simulation has more small-scale structure than the smaller box size simulations. This thus serves as a test of the impact of nonlinearities on our results. 
While this cut strictly removes small-scale modes from the initial data, any modes 
{with wavelength $\lambda<\lambda_{\rm min}$} can (and do) form once the simulation becomes nonlinear; however, the power 
{of these modes} is drastically reduced and thus so is the associated numerical error.

\subsection{Dipole cosmography prediction in post-processing}
\label{sec:dipoleprediction}

We are interested in assessing the ability of the average cosmography prediction \eqref{eqs:dLidip_av} to capture the dipole in the `true' luminosity distance as a function of redshift. We define the `true' distance in the simulation as that calculated via the fully nonlinear ray-tracing method discussed in Section~\ref{sec:rt} below. Here we detail our method for calculating the average predicted dipole by post-processing of the simulation data.

We choose a set of 20 observers placed at random locations on the final spatial slice of the simulation, which are the same observers as in the ray tracing calculations. Next these observers must also have a set of directions on their sky, namely $e^\mu$ in \eqref{eq:dLdip_predic}. We choose each direction to coincide with the center of a \texttt{HEALPix}\footnote{\url{https://healpix.sourceforge.io}} \citep{Gorski:2005} pixel for a discretised sky with $N_{\rm side}=32$.

For the predicted dipole, we need to smooth the scalar expansion rate $\theta$ and the matter density field $\rho$ on some {chosen} spatial scale. We smooth these quantities using a Gaussian kernel with a 3--$\sigma$ width that coincides with the {redshift of interest (in Appendix~\ref{appx:smoothscale} we show that this is the optimal choice for matching smoothing scale with redshift)}. {For example}, {we define the average expansion scalar as} %
\begin{equation}\label{eq:avgdef}
    \langle\theta\rangle = \sum_{a,b,c=1}^N \tilde{W}_G(x_a,y_b,z_c)\, \theta(x_a,y_b,z_c)
\end{equation}
where {$(x_a,y_b,z_c)$ is a position on the cubic grid %
and $\tilde{W}_G=W_G / \sum_{a,b,c} W_G(x_a,y_b,z_c)$} is the normalised Gaussian window function, {where}
\begin{equation}\label{eq:Gaussian_kernel}
    W_G(x,y,z) = \frac{1}{\sigma^3\, (2\pi)^{3/2}}\, {\rm exp}\left[ - \frac{(x-x_o)^2 + (y-y_o)^2 + (z-z_o)^2}{2\sigma^2}\right].
\end{equation}
{Here,} $\sigma$ is the Gaussian width (here chosen to be equal in each dimension), and the position $(x_o,y_o,z_o)$ is the location of the peak (expected value){---which} coincides with the location at which we wish to smooth $\theta$.

In {the predicted dipole} \eqref{eqs:dLidip_av}, we have terms proportional to both the smoothed $\theta$ and $\rho$ as well as their derivatives, e.g., $D_\mu \langle\theta\rangle$. For the former, we simply set $(x_o,y_o,z_o)$ to the observer position in the above formula to calculate, e.g., $\langle\theta\rangle$ {for that observer}. The values of $\theta$ at each grid cell are calculated using the \texttt{mescaline}\footnote{\url{https://github.com/hayleyjm/mescaline-1.0}} code for post-processing ET data for cosmology \citep{Macpherson:2019a} and the rest-mass density $\rho$ is output from the ET directly. 
For the derivatives, we use fourth-order finite difference approximations (in both space and time) and smooth $\theta$ and $\rho$ to each position in the required stencil. Derivatives are then taken using the smoothed values at each point.

We note that we have not included variation in the geometric volume measure of the simulation in this smoothing procedure. This would incorporate a factor of $\sqrt{\gamma}$, where $\gamma$ is the determinant of the spatial metric, into the sum \eqref{eq:avgdef}. In the simulation set-up we use, fluctuations in $\gamma$ at the grid cell scale are $\sigma_{\Delta x}(\gamma)\approx 5\times 10^{-5}$ on the final-time slice and thus represent a negligible correction to the average. 

We define the smoothing scale ({full width half maximum} of the Gaussian) as a certain number of grid cells in each simulation which we then translate into an approximate co-moving distance using the physical length of each grid cell, i.e. $L/N$ {in $h^{-1}$ Mpc}. We translate this distance {into a redshift} using the EdS distance-redshift law, with Hubble constant $H_0\rightarrow \mathcal{H}_\mathcal{D} \equiv \frac{1}{3}\langle\theta\rangle_\mathcal{D}$ where the average $\langle\rangle_\mathcal{D}$ is the fluid-intrinsic average \citep[prescribed in Section~4 of][]{Buchert:2020a} and the domain $\mathcal{D}$ is the entire simulation domain. 
{Differences in the \textit{actual} measured co-moving distance on the simulation hypersurfaces from the EdS distance are expected to be at the level of fluctuations in $\gamma$, which, as mentioned above are $\mathcal{O}(10^{-5})$ and thus below the level of numerical error.}

\section{Exact luminosity distances and redshifts}
\label{sec:exact}
In Section~\ref{sec:rt}, we review the ray-tracing method used in this work for calculating redshifts and distances between emitters and observers in the simulation. 
In Section~\ref{sec:dipDL}, we review how we extract the dipole from the sky-map of the luminosity distance of a given observer in the simulation to compare with our prediction.

\subsection{Ray-tracing method}
\label{sec:rt}

Here we provide some brief details of the ray-tracing techniques used in this work. For more details, and for tests of the code we use, see \citet{Macpherson:RT}. The ray tracer is built on the \mesc\ code \citep{Macpherson:2018akp}\footnote{Note the ray tracing part of \texttt{mescaline} is not yet public, though will be made available in the near future.} and calculates the redshift and angular diameter distance along a geodesic, with no assumptions of a particular physical form of the metric tensor (i.e., beyond gauge choices). 

The redshift is calculated by evolving the photon energy along the path of a geodesic back in time from the observer position. Specifically, we solve the geodesic equation for the photon 4--momentum $k^\mu$, %
\begin{align}\label{eq:geodesic}
    \frac{dk^\mu}{d\lambda} + \Gamma^\mu_{\alp\beta}k^\alp k^\beta = 0,
\end{align}
{where %
$\Gamma^\mu_{\alp\beta}$} are the Christoffel symbols associated with the metric tensor of space-time, $g_{\mu\nu}$. 
The angular diameter distance is defined as $D_A^2 \equiv A_s/\Omega_o$, where $A_s$ is the physical area of the source and $\Omega_o$ is its angular extent on the observer's sky. To calculate this distance we therefore must propagate an infinitesimal bundle of null rays{, rather than a single geodesic,} and track the morphology of this bundle as it traverses the space-time geometry given by $g_{\mu\nu}$.
We track the deformation of the beam by projecting it into a 2--dimensional ``screen space'' denoted by the Sachs basis vectors $s^\mu_A$ (for $A\in [1,2]$) {which is orthogonal to the direction of propagation of the bundle}. The separation of individual geodesics in the bundle %
{on the screen} evolves according to the {\textit{geodesic deviation}} equation. 
{The Jacobi matrix, $\mathcal{D}^A_{\ph{A}B}$, relates the {physical} morphology of the bundle at the source to how it appears on an observer's sky.} 
It evolves along the geodesic according to \citep[see, e.g.][]{Fleury:2015}\footnote{The positioning of the indices $A,B$ does not matter since they are raised and lowered using $\delta^A_B$, however, repeated indices still imply summation.}
\begin{equation}\label{eq:jacobis}
	\frac{d^2}{d\lambda^2} \mathcal{D}^A_{\ph{A}B} = \mathcal{R}^A_{\ph{A}C} \mathcal{D}^C_{\ph{C}B},
\end{equation}
{where} the optical tidal matrix is
\begin{equation}
	\mathcal{R}_{AB} = 
	\begin{pmatrix}
	\mathscr{R} & 0 \\
	0 & \mathscr{R}
	\end{pmatrix}
	+
	\begin{pmatrix}
	-{\rm Re}(\mathscr{W}) & {\rm Im}(\mathscr{W})\\
	{\rm Im}(\mathscr{W}) & {\rm Re}(\mathscr{W})
	\end{pmatrix},
\end{equation}
in which the Ricci and Weyl lensing scalars are {defined as}
\begin{align}
	\mathscr{R} &\equiv -\frac{1}{2} R_{\mu\nu}k^\mu k^\nu,\\ %
	\mathscr{W} &\equiv -\frac{1}{2} C_{\mu\nu\alp\beta} \sigma^\mu k^\nu k^\alp \sigma^\beta, \label{eq:curlyW}
\end{align}
respectively, where $C_{\mu\nu\alp\beta}$ is the Weyl tensor %
and $\sigma^\mu\equiv s^\mu_1 - {\rm i} s^\mu_2$ (with ${\rm i}^2=-1$).
The angular diameter distance is then the determinant of the Jacobi matrix{, i.e.,}
\begin{equation}
D_A^2 = {\rm det}|E_o \mathcal{D}|,    
\end{equation}
where $E_o$ is the energy of the photon at the observer. {We note that here the subscript $A$ in $D_A$ is solely to denote the \textit{angular} diameter distance and is distinct from the indices of the 2--dimensional screen space.}
The luminosity distance, $D_L$, is then related to the angular diameter distance via {Etherington's reciprocity relation} $D_L=(1+z)^2 D_A$. 
We refer the reader to \citet{Macpherson:RT} for technical details on the {numerical} evolution of the above system of equations using \mesc, including the calculation of the Weyl tensor, initial data, and code tests. 

In this work, as {for the dipole prediction calculation outlined} in Section~\ref{sec:dipoleprediction}, we place 20 observers at random locations within each simulation domain. The observers are all co-moving with the fluid flow, which in practice means that their 4--velocity coincides with the fluid 4--velocity vector at their location in the simulation {grid}. We propagate outgoing null rays from each observer initialized in the directions of 
\texttt{HEALPix} indices with $N_{\rm side}=32$.

\subsection{Dipole from ray-traced distances}\label{sec:dipDL}

Once the ray tracing is complete, we obtain redshifts, $z$, and luminosity distances, $D_L$, along a set of 12,288 geodesics for each of our 20 observers in each simulation. 
Each point on each geodesic lives on a sphere of constant simulation time %
surrounding the observer. This {time labeling} is an arbitrary choice and $z$ and $D_L$ thus both fluctuate across this sphere. To arrive at a physically-defined surface, we interpolate $D_L(z)$ on each geodesic to obtain constant-redshift spheres surrounding the observer. The redshift of these spheres coincides with the smoothing scale of the predicted dipole as discussed in Section~\ref{sec:dipoleprediction}.

Next, we want to extract the {dipole component of $D_L$} on a given constant-redshift sphere. We use the \hp\ \texttt{pixelfunc.fit\_dipole} function to extract the dipole vector. The norm of this vector is the dipole amplitude and we normalise the vector to obtain the unit direction for each observer. 
We use this same {procedure} to obtain an amplitude and direction of the predicted dipole calculated as described in Section~\ref{sec:dipoleprediction}.

\section{Results and Discussion} \label{sec:results}

In Section~\ref{sec:results_pred_vs_rt} we compare the ray traced dipole to our prediction based on smoothing in the simulations and in Section~\ref{sec:results_pred_vs_cg} we compare the prediction to the dipole extracted from the full third-order cosmography (without smoothing) to assess the improvement we gain by adding the smoothing procedure.

\subsection{Prediction vs ray traced dipole}\label{sec:results_pred_vs_rt}

\begin{figure*}[!b]
    \vspace{2mm}
    \includegraphics[width=\textwidth]{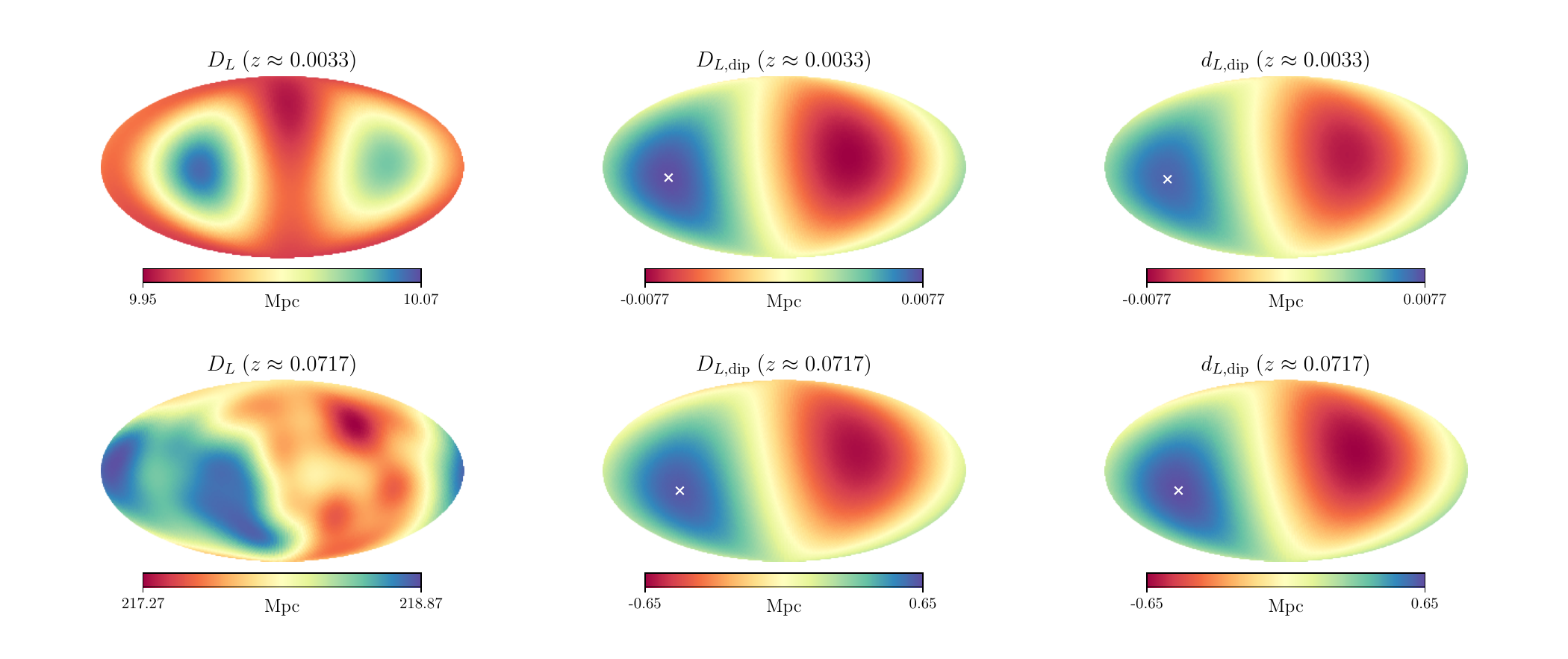}
    \caption{All-sky maps (Mollweide projection) of the ray-traced luminosity distance, $D_L$ (left panels), the dipole extracted from $D_L$ (i.e., $D_{L,{\rm dip}}$; middle panels), and the predicted dipole from the smoothed cosmography, $d_{L,{\rm dip}}$ (right panels). We show maps for one observer in the simulation with $L=2.56\,h^{-1}$ Mpc and $N=256$ on two constant-redshift slices: top row is $z\approx 0.0033$ and bottom row is $z\approx 0.0717$. For the dipoles, the fitted direction is shown with a white cross.}
    \label{fig:dip_projs}
\end{figure*}
Figure~\ref{fig:dip_projs} shows all-sky maps of luminosity distances for one observer in the $N=256$, $L=2.56\,h^{-1}$ Gpc simulation. 
{The top row shows distances for a constant-redshift slice of $z\approx0.0033$ and the bottom row for $z\approx0.0717$.} 
The left column shows the full ray-traced luminosity distance, $D_L$, %
the middle column shows the dipole extracted from {the left column}, i.e. $D_{L,{\rm dip}}$, and the right column shows the predicted dipole from the smoothed third-order cosmography, i.e. $d_{L,{\rm dip}}$. In the middle and right panels, we indicate the {fitted} dipole direction with a white cross. 
We can notice the dominance of the low-$\ell$ multipoles {in the top-left panel (i.e., at the lowest redshift) {by eye}, specifically the dipole and quadrupole.} {We see} higher-$\ell$ multipoles visible in the bottom-left panel {at slightly higher redshift}. The {predicted dipoles} in the right columns appear to {match} the ray traced dipoles in the center {columns {reasonably well} by eye}. 

We {will now} quantify exactly how well this prediction does by comparing both the amplitude and the unit direction vectors of the dipoles {for the full sample of observers}.
{In comparing the dipole direction we denote the unit dipole vector for the predicted signal as ${\bf d}$ and for the ray-traced dipole as ${\bf D}$. To determine how {closely aligned} they are to one another, we will calculate their dot product: ${\bf d}\cdot {\bf D}$.}

\newcommand{\dLdip}{d_{L,{\rm dip}}}
\newcommand{\DLdip}{D_{L,{\rm dip}}}

\begin{figure*}[t!]
    \centering
    \begin{minipage}[t]{0.48\textwidth}
    \centering
        \includegraphics[width=\textwidth]{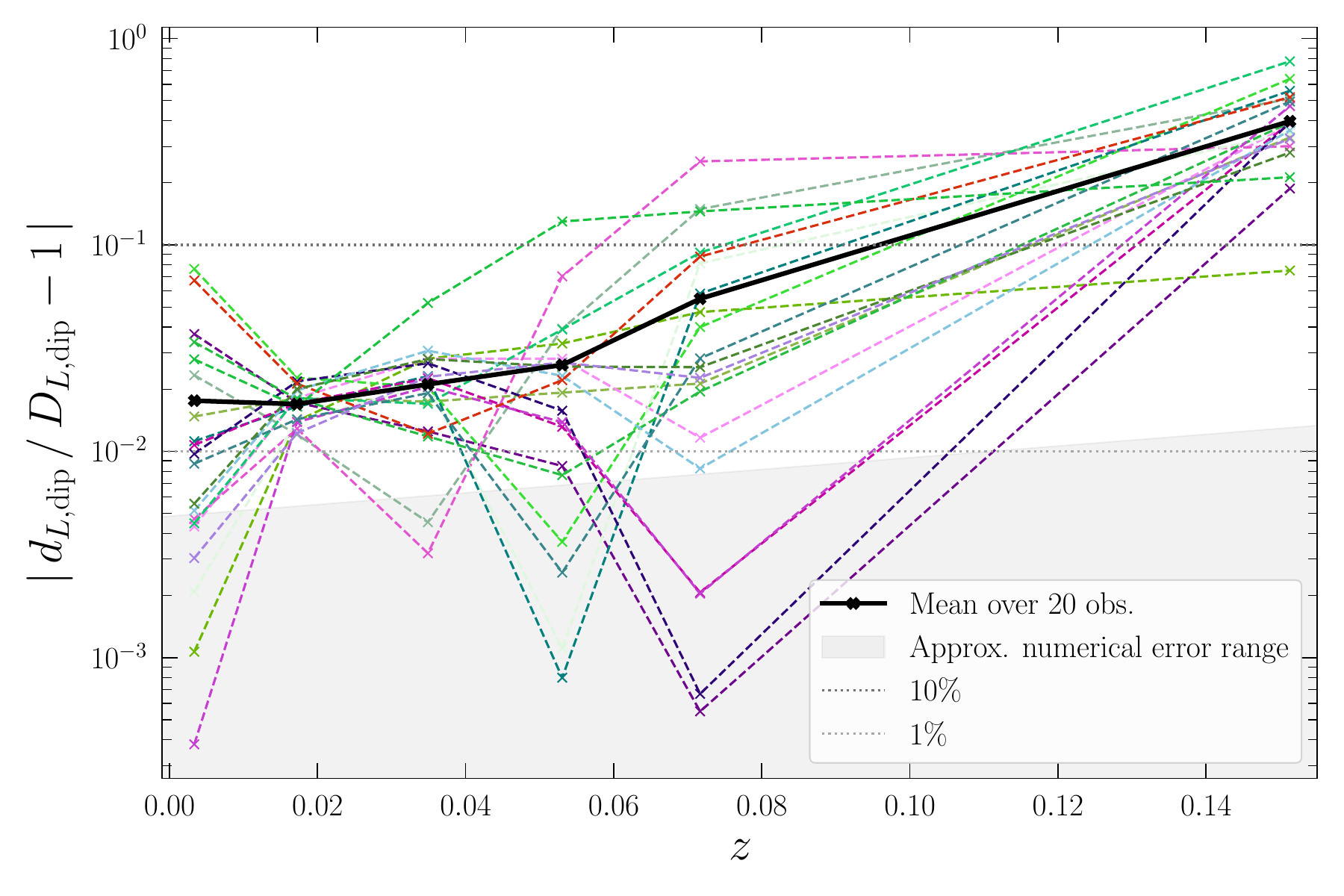}
    \end{minipage}%
    ~ 
    \begin{minipage}[t]{0.48\textwidth}
        \centering
        \includegraphics[width=\textwidth]{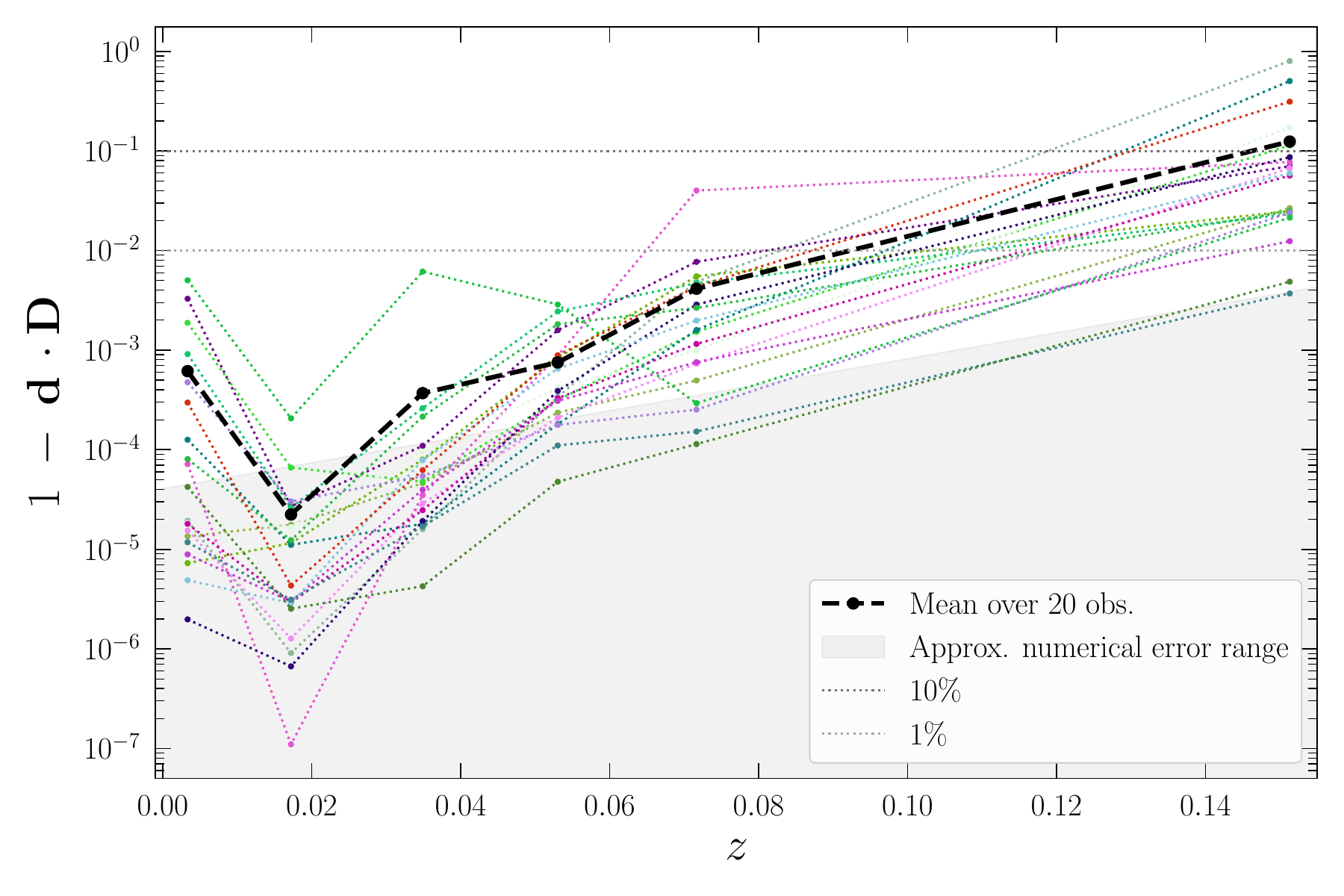}
    \end{minipage}
    \caption{Relative difference between the predicted dipole $\dLdip$ and the ray traced dipole $\DLdip$ as a function of redshift {(smoothing scale)} for a set of 20 observers in the $N=256$, $L=2.56\,h^{-1}$ Gpc simulation. The left panel shows the {absolute value of the} difference in amplitude for individual observers (dashed, coloured curves) and the average over all observers (solid, black curves). The right panel shows the alignment of the dipole unit direction vectors for the predicted signal ${\bf d}$ and for the ray-traced signal ${\bf D}$, %
    for individual observers (dashed, coloured curves) and the mean over all observers (dashed, black curve). Both panels show $1\%$ and 10\% differences as horizontal lines for reference and a shaded band representing the approximate numerical error floor for the calculations. \\}
    \label{fig:2p56_dLdip_diff}
\end{figure*}

Figure~\ref{fig:2p56_dLdip_diff} shows the relative difference between the dipole predicted 
{from the smoothed cosmography} and the ray traced dipole 
as a function of redshift for the $N=256$ $L=2.56\,h^{-1}$ Gpc simulation. The {value} on the $x$-axis corresponds to {both} the constant-redshift sphere from which $D_{L,{\rm dip}}$ is extracted {from the interpolated $D_L$} and the approximate smoothing scale used in \eqref{eqs:dLidip_av} to calculate $d_{L,{\rm dip}}$. 
The left panel shows the {absolute value of the}
difference in amplitude and the right panel shows the mismatch in dipole direction (the difference of the dot product from one). {As mentioned in Section~\ref{sec:dipcg}, the lowest-redshift point does not contain smoothed quantities and is instead the dipole prediction \eqref{eqs:dLidip} calculated directly at the grid scale.} 
Coloured curves in both panels represent the difference for individual observers and thicker black curves show the average over all 20 observers. In both panels, we show 1\% and 10\% differences as horizontal {dotted} lines for reference 
and a shaded band indicating the range $[0, {\rm max}]$ of the error as calculated across all observers\footnote{We exclude observer \#20 from the calculation of the maximum error because it has a significantly larger error bar than the full sample.}. See Appendix~\ref{appx:conv} for 
more details {on the error calculation including} error {bars} for each individual observer. {We may roughly consider the curves which fall within or below} the shaded band in Figure~\ref{fig:2p56_dLdip_diff} to be {consistent with zero} within our numerical framework. {When the difference is above this band, the mismatch in the dipoles arises from sources other than numerical error such as, e.g., higher order terms in the cosmographic expansion {which are neglected in our prediction}.}

From the left panel, we can see that the predicted amplitude matches the ray-traced dipole for all observers {at the few-percent level for $z\lesssim0.06$ and to within $\sim 10\%$ for $z<0.07$. The difference reaches $>20\%$} for most observers at the highest redshift we study here of $z\approx 0.15$. The dipole direction is matched to a higher {accuracy:} for most observers we {have} a better than 1\% prediction of the direction for $z<0.07$, and only {four} observers exceed a 10\% difference at $z\approx 0.15$. 

\begin{figure}[h!]
    \centering
    \includegraphics[width=0.6\textwidth]{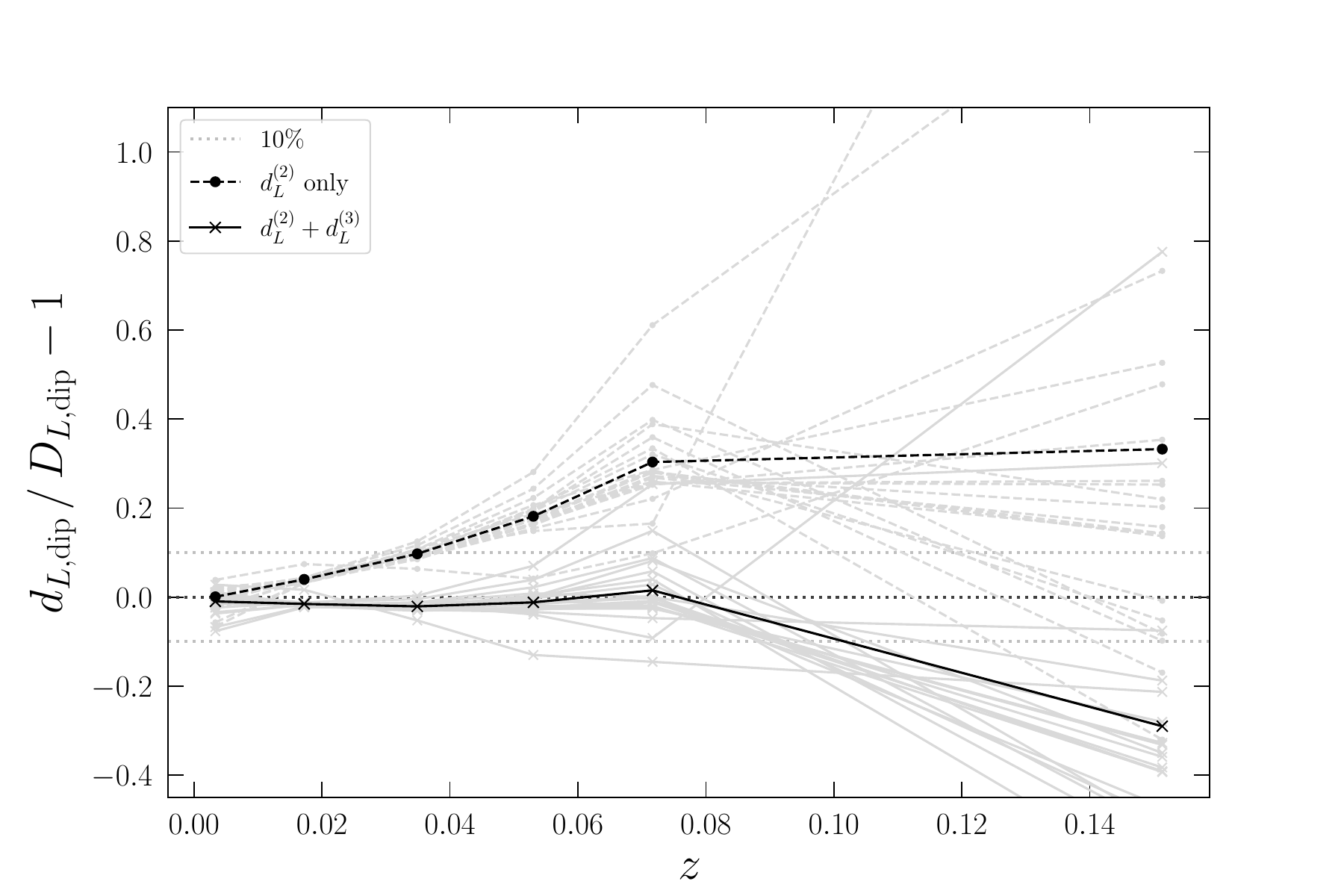}
    \caption{Relative difference between the predicted dipole $\dLdip$ and the ray traced dipole $\DLdip$ as a function of redshift. We show the difference with only the second-order contribution to the dipole prediction (i.e. only \eqref{eq:dL2dip_av}; dashed curves) and with both second- and third-order contributions (i.e., both \eqref{eq:dL2dip_av} and \eqref{eq:dL3dip_av}; solid curves). Individual observers are shown as grey curves and the mean over all observers in black. The horizontal grey dotted lines indicate a $\pm10\%$ difference.}
    \label{fig:RTvspredic_liny}
\end{figure}

While not visible in Figure~\ref{fig:2p56_dLdip_diff}, due to the absolute value on the $y$-axis, we do find a slight bias towards negative values of $\dLdip/\DLdip-1$ for the lowest redshift points. This is somewhat visible in Figure~\ref{fig:RTvspredic_liny} (solid curves{; same as those in Figure~\ref{fig:2p56_dLdip_diff}}), as well as in Figure~\ref{fig:RTvspredic_amp_werrs_3panel} in Appendix~\ref{appx:conv}. This bias appears to be due to a slight non-ideal match between the smoothing scale used in the predicted dipole \eqref{eqs:dLidip_av} and the redshift used to extract the ray-traced dipole{, which we show in Appendix~\ref{appx:smoothscale}.} 
{We thus may be able to slightly improve the power of our prediction by allowing the match between smoothing scale and redshift to vary with redshift, however, this is beyond the scope of the current work. In practise, this match could be optimised for the specific redshift range of interest in observational constraints on the dipole.}

{In these predictions so far, we have included both the second- and third-order contributions to the dipole predicted from cosmography, i.e. both \eqref{eq:dL2dip_av} and \eqref{eq:dL3dip_av}. The third-order term is more complicated in form than the second-order term, so we might consider how well the prediction fares with only the {simpler} second-order contribution \eqref{eq:dL2dip_av}.}

Figure~\ref{fig:RTvspredic_liny} shows the mismatch in the predicted dipole amplitude as a function of redshift for both second-order (dashed curves) and third-order (solid curves) accurate predictions. Individual observers are shown with grey curves and the mean over all observers is indicated with black curves. Dotted grey lines show a $\pm 10\%$ difference. The solid curves here are the same as those in the left panel of Figure~\ref{fig:2p56_dLdip_diff}, albeit without the absolute value.  
This figure demonstrates the importance of the third-order terms in making an accurate prediction {of the dipole over a broad redshift range.
While there is negligible change in the prediction for the lowest-redshift point of $z=0.0033$, we see a drastic improvement in including the third-order term \eqref{eq:dL3dip_av} in the prediction for all $z\lesssim 0.08$.}
We note that{, for this simulation, removing} the third-order contribution makes no noticeable difference to the prediction in the right panel of Figure~\ref{fig:2p56_dLdip_diff}. 
Thus, the dipole \emph{direction} is accurately captured with only the second-order contribution \eqref{eq:dL2dip_av}. This is expected, since \eqref{eq:dL3dip_av} has a contribution $\propto D_\mu \! \braket{\theta}$, which is clearly aligned with the second order term \eqref{eq:dL2dip_av}, and a contribution $\propto D_\mu \! \braket{\rho}$ which may be expected to be aligned with the direction of $D_\mu \! \braket{\theta}$ in broad classes of irrotational dust models with a Newtonian limit \citep[see equation~(17) of][]{Heinesen:2023lig}. 
However, the third-order terms are important for an accurate prediction of the dipole \emph{amplitude}.

\subsection{Comparison to local cosmography}\label{sec:results_pred_vs_cg}

Next we will compare the dipole from our prediction based on the \textit{smoothed} cosmography, i.e. \eqref{eqs:dLidip_av}, to the dipole extracted from the \textit{local} third-order cosmography \eqref{eq:series} {without any kind of smoothing}. Due to the limited radius of convergence of the generalised cosmography description \citep{Macpherson:2021gbh,Macpherson:RT}, we expect the dipole extracted from this description to perform worse than our smoothed prediction developed here. This is exactly the motivation for performing the smoothing as described in Section~\ref{sec:dipoleprediction}. We first calculate the coefficients \eqref{eq:paramseff} in the simulation at each observer position {using \mesc\ } \citep[see][for details on these calculations]{Macpherson:2021gbh} and put these into the series expansion \eqref{eq:series} for the same constant-redshift spheres as we used in Section~\ref{sec:results_pred_vs_rt} to find the luminosity distance, $d_L$, across the sky. We then extract the dipole {using \texttt{healpy} as described} in Section~\ref{sec:dipDL}.

\begin{figure*}[h]
    \centering
    \includegraphics[width=0.6\textwidth]{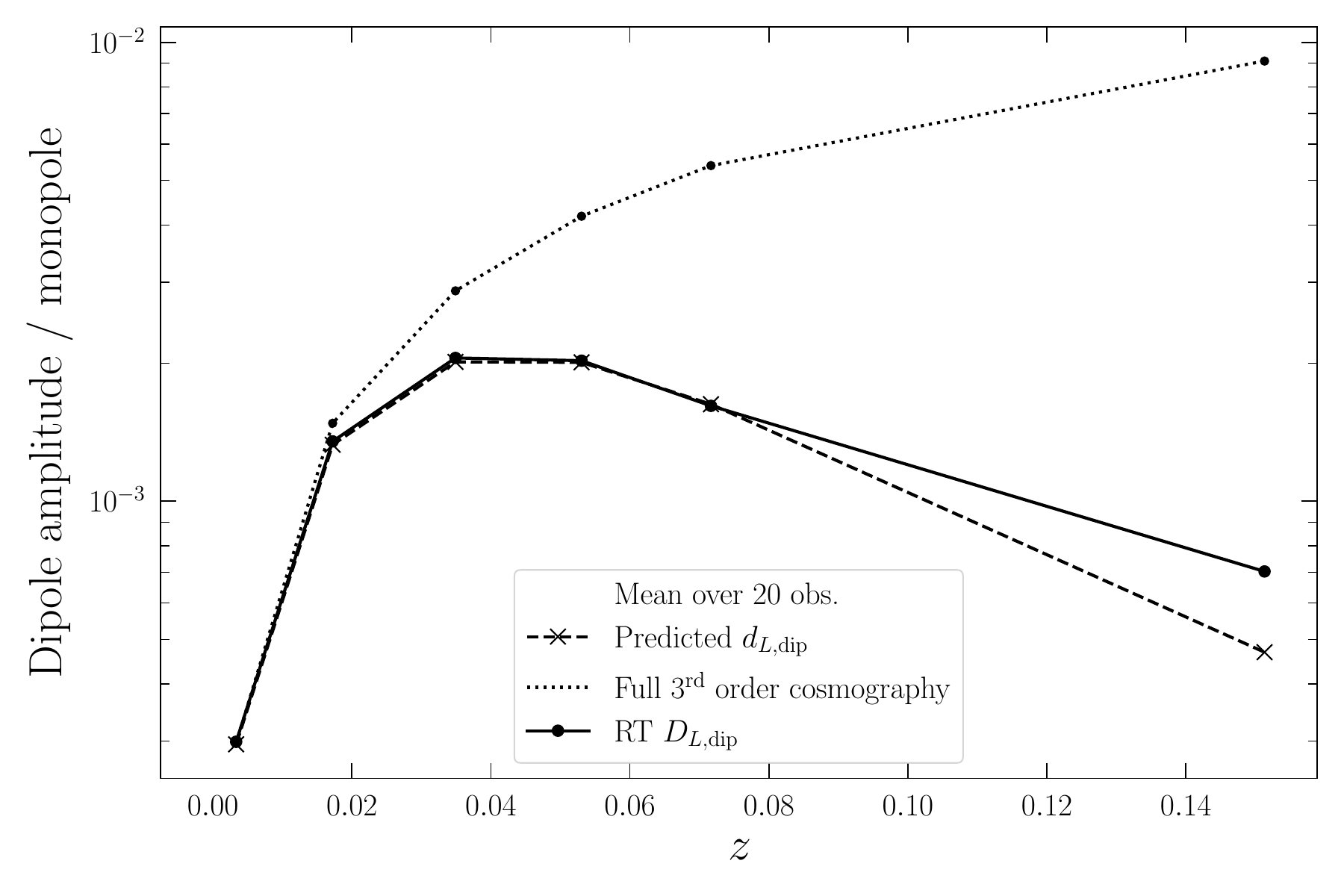}
    \caption{Dipole amplitude (normalised by the ray-traced monopole distance) as a function of redshift (smoothing scale) for the ray-traced dipole (solid curve), the predicted dipole from smoothing (dashed curve), and the dipole extracted from the local third-order cosmography (dotted curve). Each curve shows the average dipole amplitude across 20 observers. %
    }
    \label{fig:2p56_dLdip_amp_vs_z}
\end{figure*}
Figure~\ref{fig:2p56_dLdip_amp_vs_z} shows the dipole amplitude, normalised by the ray-traced monopole distance, as a function of redshift (smoothing scale for the prediction). We show the amplitude for the {`true'} dipole extracted from the ray-traced distance (solid curve), the predicted dipole amplitude via smoothing (dashed curve), and the dipole extracted from the local third-order cosmography distance (dotted curve). Each curve shows the average amplitude over 20 observers in the $N=256$, $L=2.56\,h^{-1}$ Gpc simulation. {As expected, we see an improvement in the predicted dipole by incorporating smoothing of the small-scale structures in the model universe. This improvement reaches about an order of magnitude at the highest redshift of $z\approx 0.15$}

\begin{figure*}[t]
    \centering
    \begin{minipage}[t]{0.48\textwidth}
        \centering
        \includegraphics[width=\textwidth]{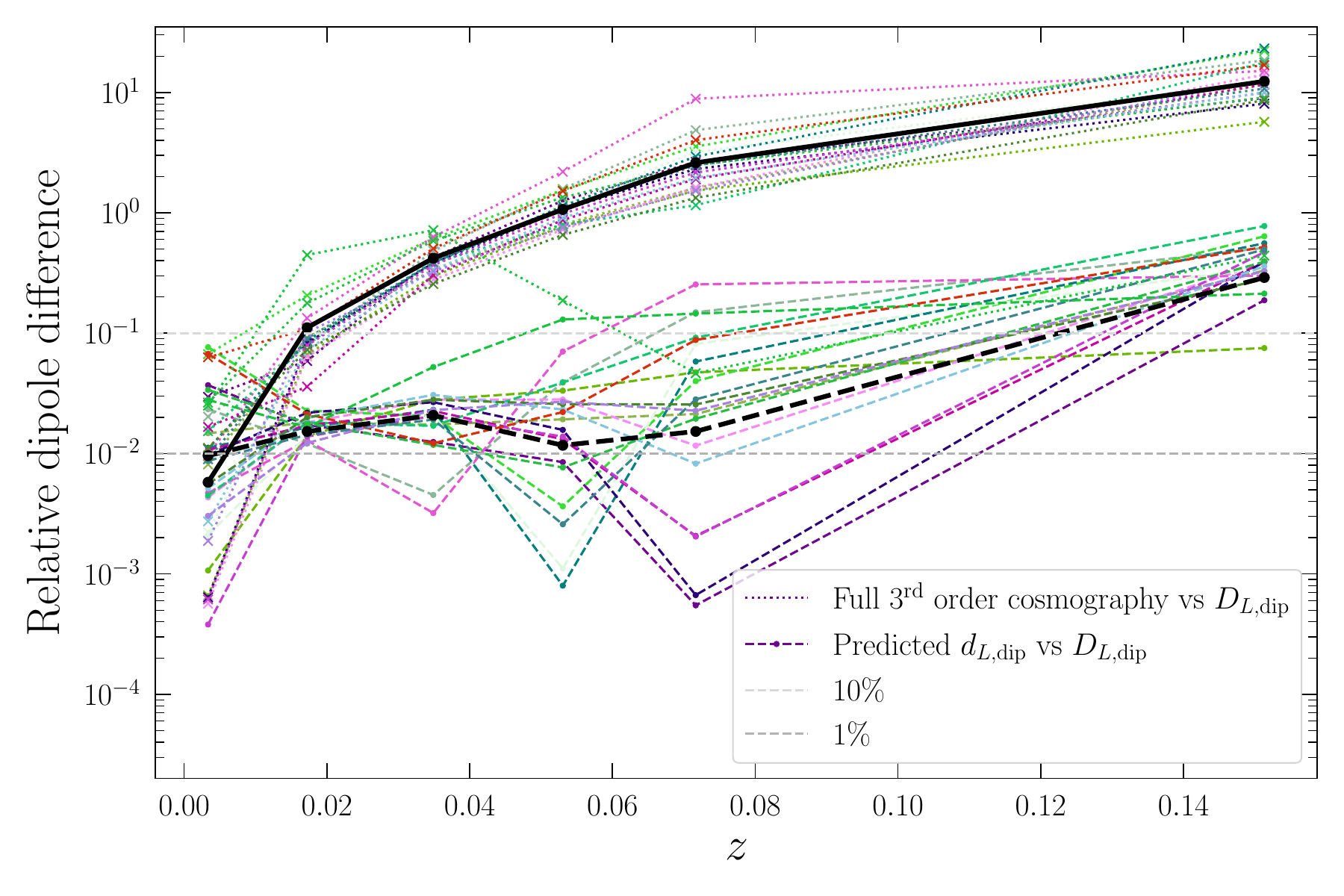}
        
    \end{minipage}%
    ~ 
    \begin{minipage}[t]{0.48\textwidth}
        \centering
        \includegraphics[width=\textwidth]{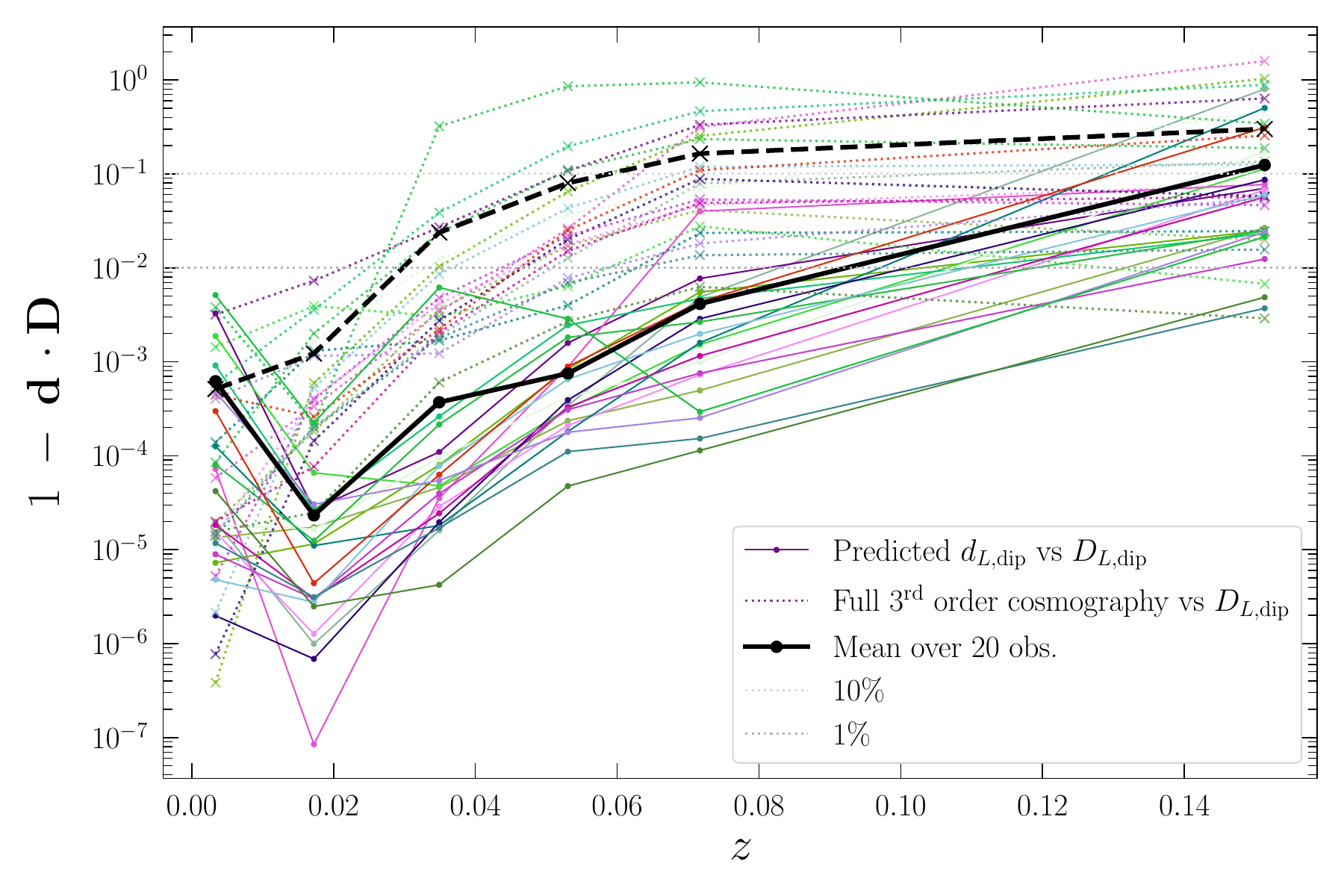}
    \end{minipage}
    \caption{Left: absolute value of the relative difference in the dipole amplitude (with respect to the ray traced distance) as a function of redshift for the smoothed prediction (dashed) and the local third order comsography (dotted). Coloured curves show the difference for individual observers and the black thick curves are the average over 20 observers. Right: difference of the dot product between the predicted and ray-traced dipole direction vectors from one as a function of redshift. Solid curves show the case where ${\bf d}$ is obtained from the smoothed prediction and dotted curves are when ${\bf d}$ is from the local third order cosmography. Black curves show the mean over 20 observers for both cases. Both panels are for observers in the $N=256$, $L=2.56\,h^{-1}$ Gpc simulation.\\}
    \label{fig:2p56_dLdip_diff_wcg}
\end{figure*}
{Figure~\ref{fig:2p56_dLdip_diff_wcg} shows the improvement in the predicted dipole amplitude (left panel) and direction (right panel) when incorporating smoothing into the generalised cosmography.}
{In the left panel, we show the difference in amplitude} with respect to the ray-traced dipole for the prediction based on smoothing (dashed curves) and for the local cosmographic distance (dotted curves). {Coloured curves show the differences for individual observers and thicker black curves are the mean over all 20 observers. The left panel of Figure~\ref{fig:2p56_dLdip_diff_wcg} quantifies the difference between the dashed and dotted curves from the solid curve in Figure~\ref{fig:2p56_dLdip_amp_vs_z}.} 
Here we see the same 1--2 order of magnitude improvement when incorporating smoothing into the dipole prediction. 
Again, we note that the left-most point in both panels corresponds to the dipole prediction \eqref{eqs:dLidip} calculated \textit{without} smoothing of $\theta$ or $\rho$. It is thus unsurprising that the prediction and the full local cosmography dipole are so similar at this scale. 

The right panel of Figure~\ref{fig:2p56_dLdip_diff_wcg} shows the mismatch in the dipole direction vectors for the smoothed prediction (solid curves) and for the direction extracted from the local third-order cosmography (dotted curves). {Coloured curves are individual observers and b}lack curves are the mean over 20 observers for both cases. At low redshift, we see about an order of magnitude improvement in the prediction of the dipole direction to $z\lesssim 0.08$ when using our smoothed method, and about a factor of $\sim 2$ for the highest redshift $z\approx 0.15$ that we study here. 

{In this section, we see a clear, significant improvement in ability of the generalised cosmography to predict the dipole signature when incorporating spatial smoothing over inhomogeneities at the corresponding scale. Next, we will investigate how well the smoothed prediction performs when smoothing over an even more inhomogeneous matter distribution.}

\subsection{The impact of small-scale structure}\label{sec:results_NL}

Our results so far have focused on the simulation with $L=2.56\,h^{-1}$ Gpc which has a significant reduction in small-scale structure in the initial data (for the purpose of the numerical convergence tests in Appendix~\ref{appx:conv}), with the smallest initial mode with non-zero power being $\lambda_{\rm min} = 400\,h^{-1}$ Mpc. This means that density contrasts in the simulation remain quasi-linear, {with typical density fluctuations at the grid scale of} $\sigma_{\Delta x}(\delta)\approx0.01$ {(and maximum values up to $|\delta|\sim 0.05$)} on the final slice {(i.e., the slice where observers are placed)}. 

We want to also test our prediction in the presence of larger density contrasts which are closer to those in the late-time Universe on large scales. While we cannot access very small scales (below $\sim 8\,h^{-1}$ Mpc) due to the limitations of our continuous fluid approximation (as discussed in Section~\ref{sec:sims}), we can reduce the initial minimum wavelength to 10 grid cells.  
We will next turn to studying the accuracy of our dipole prediction in the simulation with $N=256$ and $L=3.072\,h^{-1}$ Gpc and initial minimum mode $\lambda_{\rm min}=120\,h^{-1}$ Mpc; which at the final slice instead yields typical density contrast of $\sigma_{\Delta x}(\delta)\approx 0.05$ (and maximum values up to $|\delta| \sim 0.3$). 

\begin{figure*}[!b]
    \centering
    \begin{minipage}[t]{0.48\textwidth}
        \centering
        \includegraphics[width=\textwidth]{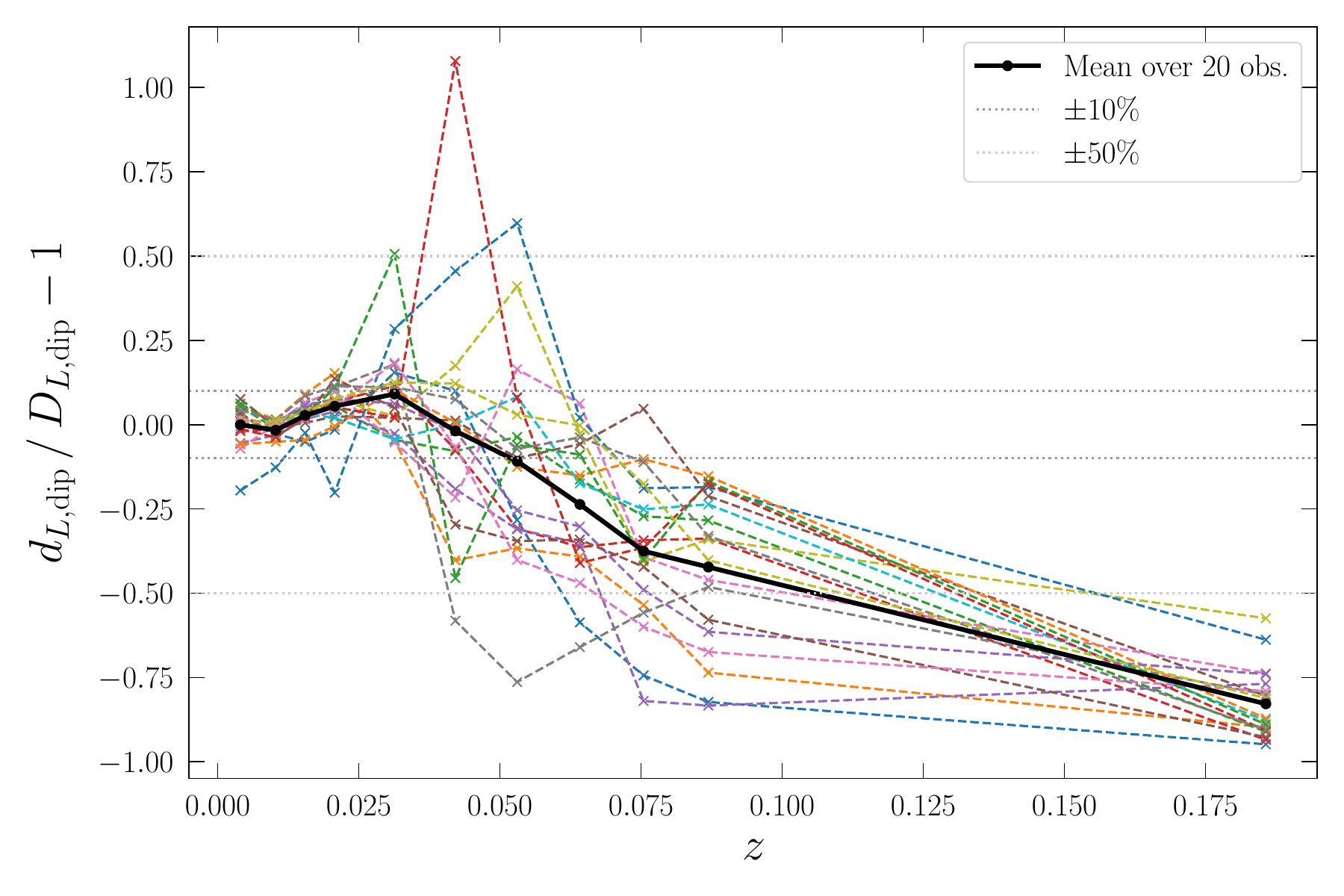}
    \end{minipage}%
    ~ 
    \begin{minipage}[t]{0.48\textwidth}
        \centering
        \includegraphics[width=\textwidth]{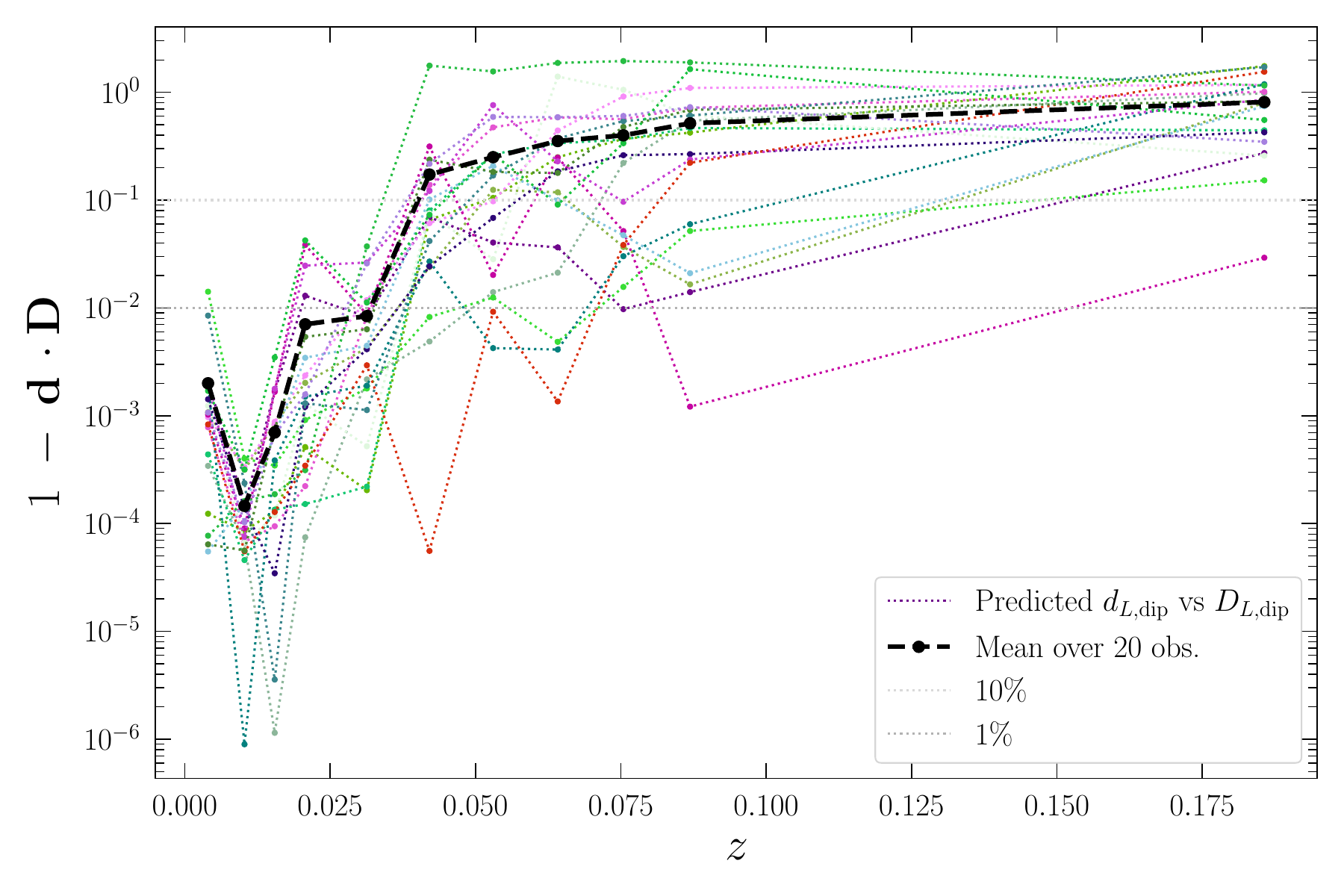}
    \end{minipage}
    \caption{Relative difference between the predicted dipole $\dLdip$ and the ray traced dipole $\DLdip$ as a function of redshift (smoothing scale) for a set of 20 observers in the $N=256$, $L=3.072\,h^{-1}$ Gpc simulation. Left panel shows the difference in amplitude for each observer (dashed, coloured lines) and for the average over all observers (solid, black line). 
    The right panel shows the alignment of the dipole unit direction vectors for the predicted signal ${\bf d}$ and for the ray-traced signal ${\bf D}$, for all observers (dashed, coloured lines) and the mean over all observers (dashed, black line). Both panels show $1\%$ and 10\% differences as horizontal lines for reference. 
    }
    \label{fig:3p072_dLdip_diff_NL}
\end{figure*}
Figure~\ref{fig:3p072_dLdip_diff_NL} shows the relative difference between the predicted dipole amplitude (left panel) and direction (right panel) and the ray traced dipole. In both panels, coloured curves show the difference for individual observers and the black curve shows the average over 20 observers. The left panel shows the %
relative difference in amplitude as a function of redshift, and the right panel shows the difference of the dot product of the unit direction vectors from one. 
Here we see the difference in amplitude rises above 10\% for most observers earlier than in the simulation in Figure~\ref{fig:2p56_dLdip_diff}. Specifically, most observers have a $\lesssim 10\%$ accurate prediction to $z\approx 0.02$, rather than $z\approx 0.07$ for the smoother simulation. We see a similar trend in the dipole direction, with the dipole vectors matching in direction to $\lesssim 1\%$ for most observers below $z\approx 0.02$ and surpassing 10\% for most observers by the highest redshift of $z\approx 0.18$ (which is slightly higher for this simulation compared to the $L=2.56\,h^{-1}$ Gpc simulation due to the slightly larger grid cell size).

We thus can see the difficulty in constructing a reliable prediction of the dipole in the luminosity distance explicitly in terms of averages of local fields over a broad redshift range in the presence of mild nonlinearities. 
This problem of construction amounts to the issue of constructing a \sayy{best fit} average 
congruence description from a local space-time description, which is at the heart of the \emph{fitting problem} in cosmology \citep{Ellis:1987zz}.
{In cases where a separation of scales is reasonable, we} do expect that our cosmography prediction {will be accurate} for a suitable average congruence description. 
{The} inaccuracy in our dipole prediction in the nonlinear case {(in Figure~\ref{fig:3p072_dLdip_diff_NL})} is likely due to the gradients of our naive average $\braket{\theta}$ not fully capturing the kinematics of the average congruence on the relevant scale.
A hopeful avenue is to develop a more accurate smoothing method to construct a suitable large-scale congruence {than we have done here, i.e.} along avenues of %
\citet{Buchert:2000,Sanghai:2015wia}, potentially tailored specifically for a particular dataset or survey redshift range. 
In the analysis of actual data, it is also possible to simply leave the construction of the underlying congruence description unspecified and \emph{assume} that a meaningful large scale congruence may be defined. 

An analysis in this spirit was performed in \citet{Adamek:2024hme} using relativistic simulations performed with \texttt{gevolution} \citep{Adamek:2016} alongside the generalised cosmographic framework. In this work, the `effective' smoothing scale is coincident with the redshift range of the survey data. This work focused on anisotropies in \lcdm\, simulations while our NR simulations do not adopt a background cosmology. In principle \texttt{gevolution} simulations could be used to study the accuracy of our dipole prediction in \lcdm, although we expect similar issues in matching an average congruence description in a nonlinear simulated universe.

\subsection{Convergence to large-scale average frame}

For both simulations, we find that the dipole prediction, $\dLdip$, under-predicts the ray-traced dipole $\DLdip$ at the highest redshifts we study (i.e., the negative differences in the solid curves of Figure~\ref{fig:RTvspredic_liny} and the left panel of Figure~\ref{fig:3p072_dLdip_diff_NL}). 
The failure in this regime is somewhat expected because the {dipole prediction from the average} cosmography is failing to account for local kinematics below the scale of averaging. 
While the local kinematics is largely expected to cancel out across large averaging scales, the  
motion of the observer below the smoothing scale is a systematic that creates a dipole in addition to that predicted by the large-scale cosmography.  
At large scales, the gradients inducing the cosmography dipole in \eqref{eqs:dLidip_av} will be greatly reduced in amplitude and the dominant contribution will instead be caused by the observer's peculiar motion with respect to the averaged frame. 
We thus expect that $\DLdip$ will be well described at these scales by a dipole induced by a special-relativistic boost of the observer.

In this section, we will test this expectation by calculating the boost expected for our observers from their velocity with respect to the large-scale average frame at a given scale. We define the boost velocity as
\begin{equation}
    v^i_B \equiv v^i - \langle v^i \rangle,
\end{equation}
where $v^i\equiv u^i / (\alpha u^0)$ is the observers peculiar velocity with respect to the simulation hypersurface frame and $\langle v^i\rangle$ is $v^i$ smoothed in the same way as described in Section~\ref{sec:dipoleprediction}.
 
The dipole in the luminosity distance induced by this boost---where we assume the large-scale average frame is well-described by the EdS model---is \citep{Bonvin:2006en} 
\begin{equation}\label{eq:dipboost}
    d_{L,{\rm boost}}(z) = \frac{(1+z)}{\mathcal{H}(z)} v_B^i e_i ,
\end{equation}
where $e^i$ is the direction vector of the incoming geodesic and $\mathcal{H}(z)$ is the EdS Hubble parameter at the given redshift, which we calculate (in conformal time) using $\mathcal{H}(z)=\mathcal{H}_0 \sqrt{1+z}$, where $\mathcal{H}_0=\langle\theta\rangle/3$ for each observer at the scale of interest. Since we expect the convergence of $\DLdip$ to $d_{L,{\rm boost}}$ to occur at large scales, we include additional smoothing scales with respect to 
earlier figures to a redshift of $z\approx 0.4$. 

\begin{figure*}[!t]
    \centering
    \includegraphics[width=0.6\textwidth]{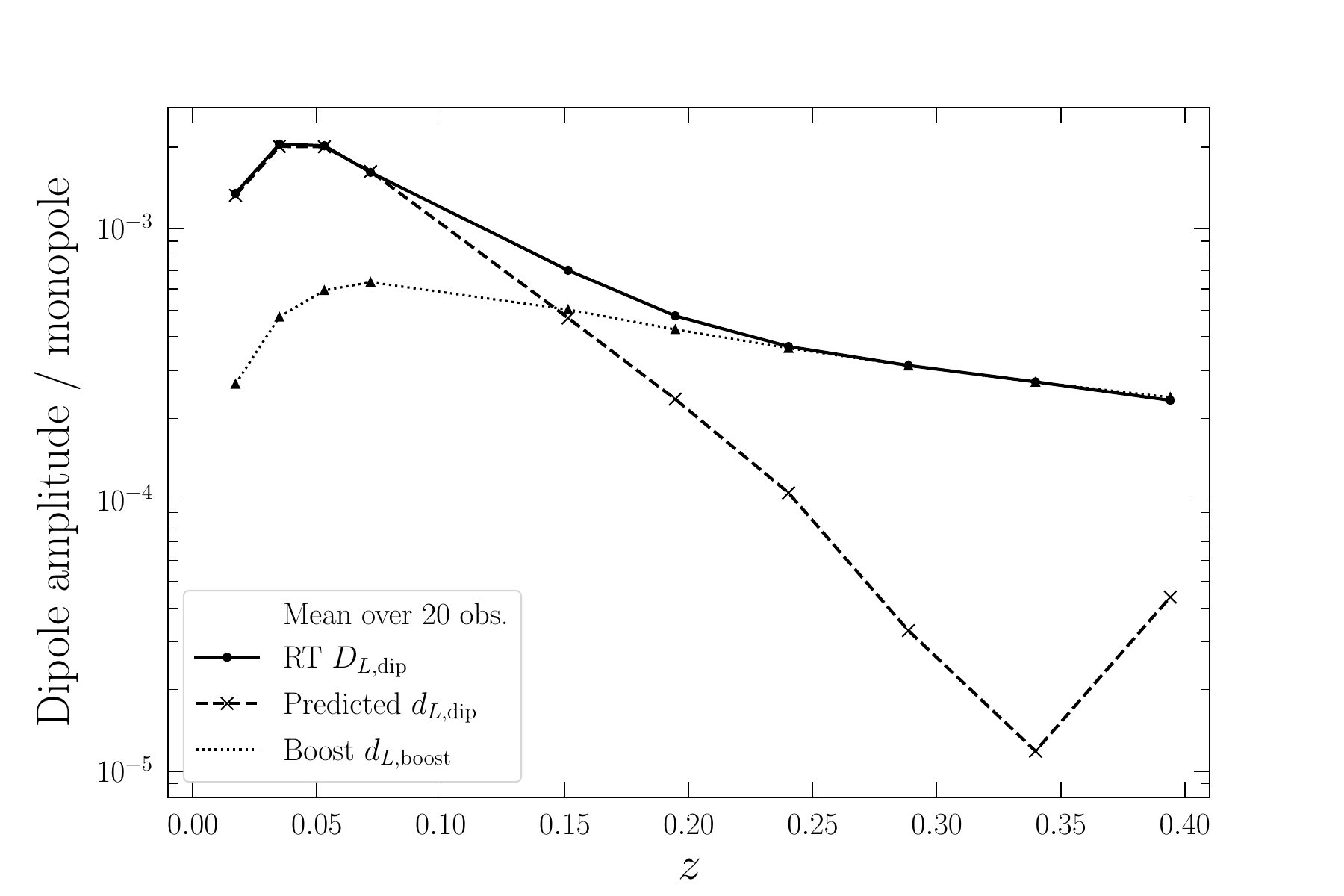}
    \caption{Dipole amplitude (normalised by the ray-traced monopole distance) as a function of redshift (smoothing scale) for the ray-traced dipole (solid curve), the predicted dipole from smoothing (dashed curve with crosses), and the dipole from a single boost of the observer (dotted curve with triangles). Each curve shows the average dipole amplitude across 20 observers for the simulation with $N=256$ and $L=2.56\,h^{-1}$ Gpc.
    }
    \label{fig:2p56_dLdip_amp_vs_z_boost}
\end{figure*} 
 
Figure~\ref{fig:2p56_dLdip_amp_vs_z_boost} shows the dipole amplitude as a function of redshift/smoothing scale normalised by the monopole ray-traced distance. We show the ray traced dipole (solid curve), the predicted dipole $\dLdip$ (dashed curve with crosses), and the dipole purely from a boost (via \eqref{eq:dipboost}; dotted curve with triangles). 
Each curve shows the average dipole amplitude across all 20 observers in the $L=2.56\,h^{-1}$ Gpc simulation with $N=256$. We see the smoothed cosmography prediction gives the best match at low redshifts of $z\lesssim 0.1$, whereas the ray traced dipole converges towards the dipole expected from a boost at higher redshifts of $z\gtrsim 0.25$. 

Within a perturbative framework, the ray traced dipole can be understood as containing contributions from both local gradients and peculiar velocity effects. At low redshifts, the smoothing scale is close to the scale of local kinematics and so the velocity of the observer/sources is small with respect to the smoothed frame. Thus, the contribution from local gradients is dominant and we see the ray traced dipole is best matched by the smoothed cosmography description in Figure~\ref{fig:2p56_dLdip_amp_vs_z_boost}. At high redshifts, with a larger smoothing scale the dipole contribution from local gradients is suppressed (dashed curve in Figure~\ref{fig:2p56_dLdip_amp_vs_z_boost}) and the contribution from the observer velocity with respect to the smoothed frame (dashed curve) dominates the ray traced dipole. 
We might consider whether some combination of the two might allow us to improve our prediction in the mid-range of $0.1\lesssim z \lesssim 0.25$ where we see neither the smoothed cosmography or the simple boost dipole alone provide an accurate description of the ray traced dipole. 
However, such a combined framework would likely require the use of perturbation theory to completely isolate these two effects. The adoption of perturbative frameworks is not in the spirit of the generalised cosmography and thus is beyond the scope of this paper.

\subsection{Connection to observable dipoles}

A key motivation for developing our predictive framework is for the application to observational analyses outside of common simplified models such as FLRW and/or perturbation theory. In this work, we have provided the first quantification of the ability of our prediction to stand up to dipoles in {mildly} nonlinear relativistic simulations. 
In particular, our results show that measurements of a dipole in luminosity distances in the nearby Universe can be reasonably interpreted in terms of gradients in expansion and matter density. While this interpretation is dependent upon general relativity, it can be made completely free of any particular background cosmological model (though must lie within the broad class of quiet universe cosmologies outlined in Section~\ref{sec:dipcg}). This is potentially useful for applications to, e.g., constraints on dipoles which are currently in tension with $\Lambda$CDM and may benefit from a model-independent analysis. 
{While a cosmography dipole may arise due to peculiar bulk motions in $\Lambda$CDM cosmology, it may also incorporate non-$\Lambda$CDM behavior, and anisotropic cosmography frameworks are therefore ideal for scenarios that cannot necessarily be captured completely by $\Lambda$CDM.  }

Low-redshift dipoles in the luminosity-distance redshift relation have been measured with some success in supernova catalogues \citep[with varying levels of significance, e.g.][]{Dhawan:2022lze,Cowell:2022ehf,Colin:2019a,Rubin:2020,McConville:2023,Rahman:2022,Kalbouneh:2022tfw,Sorrenti:2023}. Upcoming low-redshift supernova catalogues, such as from the Zwicky Transient Facility \citep[ZTF;][]{Rigault:2025}, are expected to contain several thousands of objects which can be used for cosmological analyses. Additionally, the Vera C. Rubin Observatory's Legacy Survey of Space and Time (LSST) is conservatively expected to discover $\sim$10,000 supernovae per year \citep{LSST:2009}. With these vast increases in numbers of objects as well as improved sky-distribution of the surveys, we will be able to improve measurements of the dipole in the luminosity-distance redshift relation significantly \citep{Dhawan:2022lze}. 

It would be an interesting avenue to explore similar cosmography methods to address number count dipoles, that are for instance {measured} through the Ellis--Baldwin test \cite{1984MNRAS.206..377E,Secrest:2025nbt}. 
Typically, number counts are investigated in an integral sense, so that one ends up with data defined on the 2--sphere. 
However, slices in redshift can be considered for the Ellis-Baldwin test \citep{Maartens:2017qoa,vonHausegger:2024fcu}, and cosmography approaches could be formulated for the low-redshift regime. This seems especially relevant in light of recent findings that suggest that the cosmic dipole as seen in the number counts may not be of kinematic nature \citep{Wagenveld:2025ewl}. If indicating an actual cosmological signal, the excess number count dipole may be expected to correlate with the dipole in the luminosity distance.   

{It would also be interesting to explore dipoles in other observables, such as that in the position drift measured through proper motions of quasars in the Gaia Celestial Reference Frame dataset \citep{gaia2021acceleration,Bel:2018} with the covariant cosmography approach outlined in \citet{Heinesen:2024npe}. %
Further into the future, studying anisotropic cosmography with redshift drift \citep{Heinesen:2021qnl,Oestreicher:2025} would also be interesting 
via the average framework presented in this paper. }

\section{Conclusions}
\label{sec:conclusion}

Our main theoretical result is {\eqref{eqs:dLidip_av}:} the prediction of the luminosity distance dipole based on the smoothed cosmography. 
We have assessed this prediction against ray-traced 
luminosity distances and redshifts within 
fully relativistic NR simulations. 
We found in quasi-linear simulations (with percent-level typical density contrasts) that the prediction accurately captures the dipole in luminosity distances within 10\% for most observers for redshifts below $z=0.07$. For a {more inhomogeneous} simulation (with few- to tens-of-percent density contrasts), the prediction is accurate to within 10\% instead for redshifts below $z=0.02$. As noted in \citet{Macpherson:2021gbh} and \citet{Macpherson:RT}, the presence of smaller-scale inhomogeneities spoils the convergence of the cosmographic relation. However, we see up to an order of magnitude improvement when incorporating smoothing into the dipole prediction {as opposed to the dipole predicted from the \textit{local} cosmography studied in \citet{Macpherson:2021gbh}.}

{A key caveat to this work is that we are limited in the amount of small-scale structure that can exist in our simulations due to the adoption of a continuous fluid approximation. This inherently means that our results will change as we change resolution and thus sample smaller, non-linear scales (as can be seen in the differences between Figure~\ref{fig:2p56_dLdip_diff} and Figure~\ref{fig:3p072_dLdip_diff_NL}). This issue can be solved by incorporating particle dynamics into our simulations which would allow for the formation of bound structures; allowing us to accurately simulate structures below our current minimum scale of $\sim 8\,h^{-1}$ Mpc. 
Smoothed particle hydrodynamics has been incorporated into the ET via the \textsc{phantom} code \citep{Magnall:2023,Price:2018}, however, more development is required for realistic, large-scale cosmological simulations.
Using these kinds of advanced simulations to assess the accuracy of our dipole prediction in the deeply nonlinear regime could prove important for its application to constraints in the late-time Universe.}

Here we have studied matter-dominated simulations (i.e., with no dark energy) which have very small backreaction effects both on spatial sections \citep{Macpherson:2018akp,Macpherson:2019a} and in observable quantities \citep{Macpherson:RT,Macpherson:2024zwu}---thus remaining very close to the EdS model evolution on large scales. 
The question remains as to how far one may extrapolate the dipole prediction given in \eqref{eqs:dLidip_av} to alternative cosmological models. 
We expect that the prediction will be similarly accurate for all universe models that are well described by the quiet universe approximation outlined in Section~\ref{sec:dipcg} and which have sufficient decoupling of scales such that the mappings $\theta \mapsto \braket{\theta}$ and $\rho \mapsto \braket{\rho}$ can be done in transitioning from the local scale to a larger scale.

A key issue we have highlighted is in applying our dipole prediction to model universes with large density contrasts. Paying closer attention to finding the ideal method by which to smooth the inhomogeneous model universe to a well-fitting, large-scale model may improve our prediction. However, this is a well-known and highly nontrivial issue in cosmology \citep[e.g.][]{Ellis:1987zz,Clarkson:2011b}. 
One way to circumvent such issues would be to write the cosmography in a frame which is \emph{not} comoving with the matter and in which the expansion scalar $\theta$ remains linearly-perturbed when such a frame may be identified.  
An example is the quasi-Newtonian class of space-times investigated in \cite{Heinesen:2023lig}, where the primary dipole may be understood as being generated by the bulk matter flow relative to the quasi-Newtonian frame assumed to exist in which the extrinsic curvature remains linearly perturbed. 
While such representations of the cosmography may be convenient to work with for the purpose of averaging, they are limited to certain classes of space-times, and they are less natural to work with in terms of observations as they require the identification of an idealized quasi-Newtonian frame, which is in practice challenging.  
It is beyond the scope of this paper to go into such alternative strategies for predicting the dipole in distance--redshift data. 

The problem of constructing large-scale observational predictions in a lumpy space-time can be seen as a manifestation of the \emph{fitting problem} in cosmology \citep{Ellis:1987zz}, where the aim is to construct a large scale congruence description valid for describing average dynamics and light propagation in a locally arbitrarily-inhomogeneous universe.
As we have remarked in the above, in the analysis of observational data it is possible to simply assume that a meaningful large scale congruence  
exists and to 
constrain it directly from distance--redshift data using a framework like ours.

\begin{acknowledgments}

We would like to thank the anonymous referee whose comments improved the clarity of our manuscript. 
Support for HJM was provided by NASA through the NASA Hubble Fellowship grant HST-HF2-51514.001-A awarded by the Space Telescope Science Institute, which is operated by the Association of Universities for Research in Astronomy, Inc., for NASA, under contract NAS5-26555. AH is supported by the Carlsberg Foundation. Simulations and post-processing analyses used in this work were performed with resources provided by the University of Chicago's Research Computing Center.

\end{acknowledgments}

\appendix 

\section{Numerical convergence tests}\label{appx:conv}

Our results presented in the main text will contain {numerical errors} due to the finite accuracy of the simulations themselves. This error is dependent on the accuracy of the  scheme used for numerical integration/differentiation (of both the simulation and the post-processing analysis), the accuracy with which the initial data satisfies the constraint equations of general relativity, and the resolution of the simulation. 
The numerical error is proportional to $(\Delta x)^n$ where $n$ is the order of the numerical scheme and the cubic grid spacing is $\Delta x = L / N$, where $L$ is the comoving side length of the total simulation domain, and $N^3$ is the number of cells. 
The evolution of our simulations in the ET uses a Runge-Kutta fourth-order (RK4) integrator, i.e. with error of order $(\Delta x)^4$. Our ray tracing analysis with \texttt{mescaline} uses RK2\footnote{There is an RK4 integrator available for ray tracing in \texttt{mescaline}, however, we use the RK2 integrator here to reduce the high computational expense of the calculations; the same which was used for the tests in \citet{Macpherson:RT}.} with error of order $(\Delta x)^2$. Our calculation of the predicted dipole adopts fourth-order accurate finite-difference derivative stencils. 
The initial data we have used here solves the constraint equations identically to the fourth-order accuracy of finite-difference derivatives on the initial slice. Based on this, we expect our ray-traced results to converge with rate $(\Delta x)^2$ and the prediction to converge with rate $(\Delta x)^4$.

To place error bars on the calculations in our main results, we will perform a Richardson extrapolation on the dipole amplitude and direction vectors. The Richardson extrapolation is based on the premise that as we increase the resolution of our simulation towards infinity our errors approach zero, i.e. $N\rightarrow\infty$ implies $\Delta x \rightarrow 0$, and our calculations represent the `true' result. Even in the limit of infinite numerical resolution, our simulations and analyses contain some specific choices which will influence the final results. For example, the initial power spectrum cut of the simulation does change the physics of the simulation and thus will change our results, similarly the choice of observers as comoving with the fluid flow (i.e., without a boost with respect to this frame) will change the dipole signal they measure in distances. Choices like these will inherently change the physical results we would obtain in this work. Thus, we refer to the `true' solution here as simply the solution we would obtain in our chosen physical set-up without the influence of numerical error. 

The Richardson extrapolation requires at least three simulations with different $N$ so that we can extrapolate to $N=\infty$ and estimate the true solution, and the error bar is then the difference between this true solution and our highest-resolution results (which are those we present in the main text). To isolate the impact of changing $\Delta x$ on our results, these three simulations \textit{must} be as physically close as possible. We use $N=64, 128$, and 256 each with identical initial data generated as follows: first we make initial data for the $N=64$ and $L=2.56\,h^{-1}$ Gpc simulation from an initial power spectrum with minimum scale of $\lambda_{\rm min} = 10 \Delta x = 400\,h^{-1}$ Mpc. We then interpolate the resulting $\phi(x^i)$ field to $N=128$ and $N=256$ points using a quintic spline interpolation (via the \texttt{SciPy.interpolate.interpn} function). These $\phi(x^i)$ are then input into \texttt{FLRWSolver} to solve the Hamiltonian and momentum constraints according to Section~\ref{sec:ICs} for each individual resolution. We use the same 20 observers with identical positions (and lines of sight) in each of the three simulations for both the ray tracing and dipole prediction analysis in post-processing.

An example of how the Richardson extrapolation works is the following: we calculate the dipole amplitude for the ray traced distances for one observer in each simulation with $N=64, 128$, and 256. We can thus plot $D_{L,{\rm dip}}(N)$ with three points and fit a curve of the form $f(N)=a+b/N^2$ where $a$ represents the `true' solution in the case $N\rightarrow\infty$ that we discussed earlier in this section and $b$ is the contribution from numerical error. Here, the power of $N$ represents the order of accuracy of the scheme: for the predicted dipole this power would change $2\rightarrow 4$. We use the \texttt{SciPy.curve\_fit} function to find $a$ and $b$ and the error is then defined as %
$D_{L,{\rm dip}}(256) - a$. Not every set of $D_{L,{\rm dip}}(N)$ points will lie perfectly on the curve $f(N)$ that we use to find the error. The main reason for this would be if an observer was in a particular location in the simulations where the physical structures around them were not consistent enough between resolutions, i.e. such that \textit{physical} changes in $D_{L,{\rm dip}}$ as we changed $N$ might be comparable to, or sometimes larger than, the numerical changes. The reason this can occur is as we increase resolution we naturally expect structures to form which are below the resolution scale of the lowest resolution simulation, which can lead to differences in e.g. the density field. This will then leak into differences in the dipole and we are thus no longer isolating the numerical error in our calculations. We expect such observers to be a minority given the care we have taken to keep the initial data the same between resolutions. To find observers who might be in such special locations, we assess the coefficient of determination, $R^2\equiv 1 - SS_{\rm res}/SS_{\rm tot}$, of the curve fit. Here $SS_{\rm res}\equiv \sum_N (D_{L,{\rm dip}}(N) - f(N))^2$ is the sum of the square of the residuals between the data and the function and $SS_{\rm tot}\equiv \sum_N (D_{L,{\rm dip}}(N) - \langle D_{L,{\rm dip}}\rangle_N)^2$ is the variance of the data. A value of $R^2=1$ indicates a perfect fit where the function lies exactly on the data points. Here, we will qualify a value of $R^2\geq 0.8$ as a `good' fit.

\begin{figure*}
    \includegraphics[width=\textwidth]{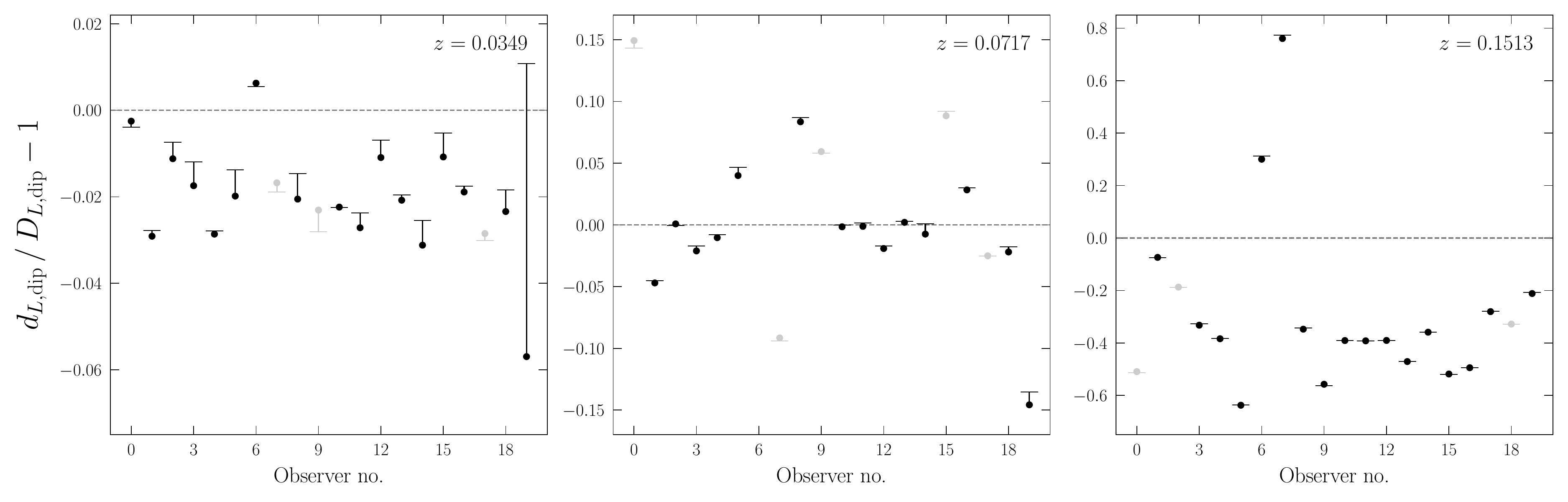}
    \caption{Relative difference in the predicted dipole amplitude with respect to the ray traced dipole for all 20 observers and for three redshifts (panels; as indicated in the legends). We show the difference for the third-order prediction only and the error bar is calculated via a Richardson extrapolation (points show simulation-calculated values and bars show Richardson-extrapolated values). 
    Grey points indicate observers with $R^2<0.8$ in the Richardson extrapolation. 
}
\vspace{4mm}
    \label{fig:RTvspredic_amp_werrs_3panel}
\end{figure*}
Figure~\ref{fig:RTvspredic_amp_werrs_3panel} shows the mismatch between the predicted dipole and the ray traced dipole for the 20 observers in the $N=256$, $L=2.56\,h^{-1}$ Gpc simulation. Each panel shows the relative difference in dipole amplitude for a different constant-redshift slice, with error bar calculated via a Richardson extrapolation as described above. The errors are calculated on the quantity $\dLdip/\DLdip-1$ directly rather than on the individual dipole amplitudes. The points indicate the actual quantity that was calculated in the $N=256$ simulation and the horizontal bar shows the Richardson extrapolated value. 
Black points indicate observers with a `good' fit in the Richardson extrapolation of $R^2\geq 0.8$ and grey points are observers with a `bad' fit with $R^2<0.8$. 
At low redshifts, the match between the predicted and ray traced dipole improves for the Richardson extrapolated value relative to the $N=256$ calculated value for most observers. This is expected, since the Richardson extrapolated value is an estimate of the idealized case  $N \rightarrow \infty$ without numerical error. 
We note that there are fewer points here than the equivalent redshift points in Figure~\ref{fig:2p56_dLdip_diff}. 
This is simply due to the fact that in order to do the Richardson extrapolation we require points at all three redshifts. The lowest resolution, $N=64$, has fewer points in redshift along the geodesics from the ray tracing analysis due to the larger grid cells in the simulation. The three redshift panels in Figure~\ref{fig:RTvspredic_amp_werrs_3panel} represent the only points for which all three resolutions have data at the same redshift.

\begin{figure*}
    \includegraphics[width=\textwidth]{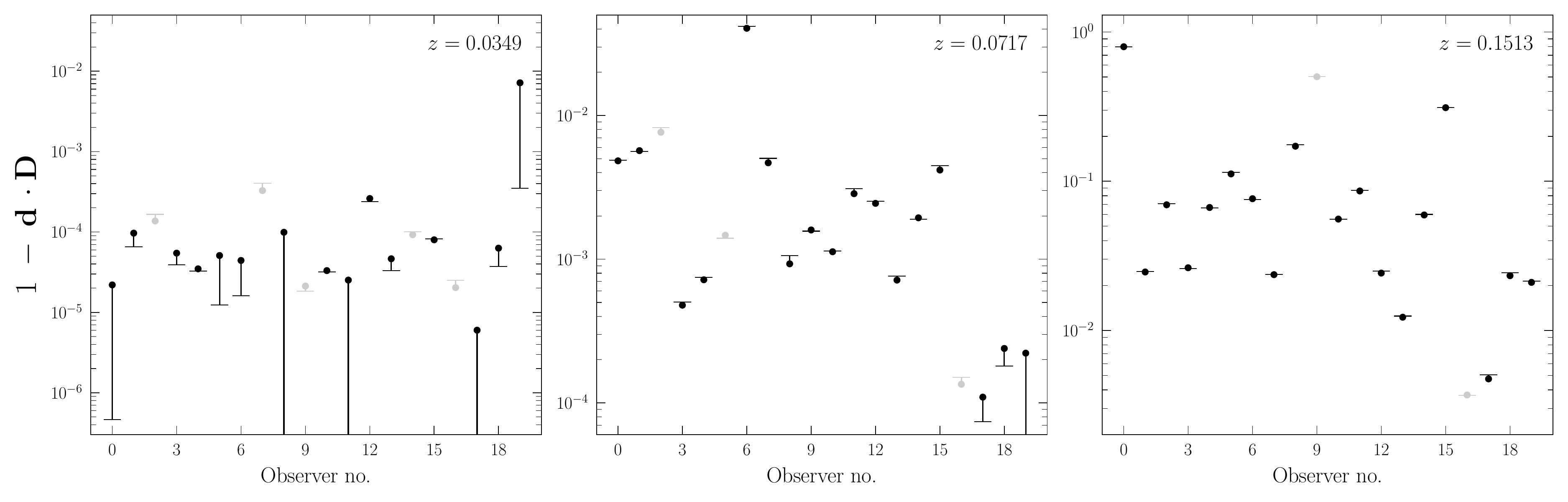}
    \caption{Difference in the dot product of the unit dipole direction vectors from one for the same 20 observers as in Figure~\ref{fig:RTvspredic_amp_werrs_3panel}. The prediction here is up to third-order terms, with no noticeable difference when using only second-order terms in the prediction. Grey points indicate observers with $R^2<0.8$ in the Richardson extrapolation. Points show simulation-calculated values and bars show Richardson-extrapolated values. }
    \label{fig:RTvspredic_vecdot_werrs_3panel}
\end{figure*}
Figure~\ref{fig:RTvspredic_vecdot_werrs_3panel} shows the mismatch in the dipole direction vector; specifically, the difference of the dot product of the predicted and ray traced unit vectors from one. The error bars on each point were calculated via a Richardson extrapolation directly on the quantity $1-{\bf d}\cdot{\bf D}$ rather than on the components of each direction vector independently. Points again indicate the value calculated in the $N=256$ simulation and the horizontal bar shows the Richardson-extrapolated value using all three resolutions. 

The main conclusion from this test is that, for most observers, the mismatch between the predicted and ray-traced dipole is not dominated by numerical error. {A few} observers at the lowest redshift show that the mismatch in direction is dominated by numerical error, and one in the amplitude. We can thus conclude that the predominant reason for the mismatch at these three redshifts is higher-order terms in the cosmography prediction that are neglected in our third-order treatment. Unfortunately, we {cannot use the Richardson extrapolation method} for the two lowest-redshift points in Figure~\ref{fig:2p56_dLdip_diff}, since we do not have data at three resolutions here. However, given the mismatch is of a similar order of magnitude to the left-most panel of Figure~\ref{fig:RTvspredic_amp_werrs_3panel} we might expect a similar level of numerical error.

\section{Matching the smoothing scale with a redshift}\label{appx:smoothscale}
\begin{figure*}
    \centering
    \begin{minipage}[b]{0.49\textwidth}
        \centering
        \includegraphics[width=\textwidth]{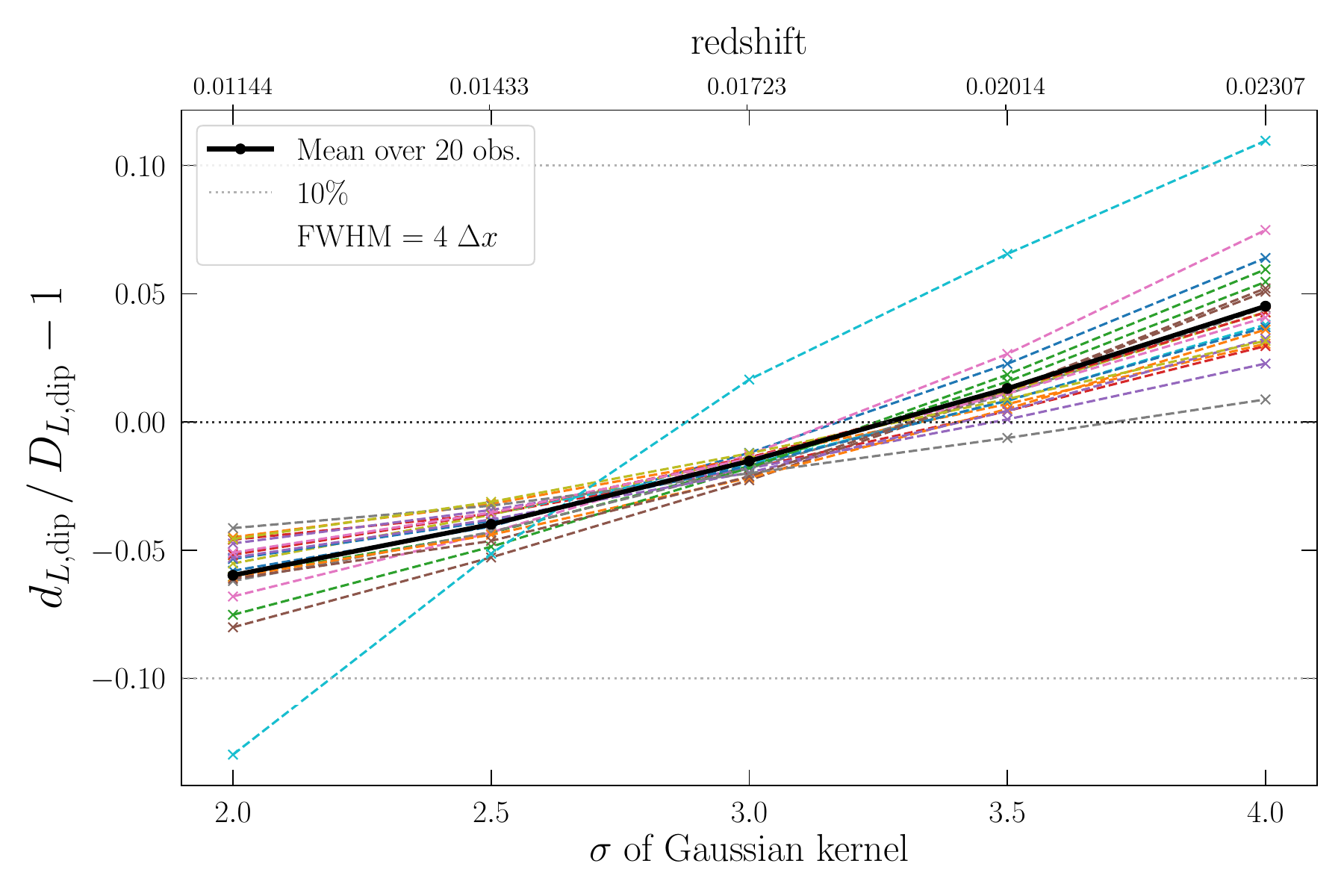}
    \end{minipage}
    \hfill
    \begin{minipage}[b]{0.49\textwidth}  
        \centering 
        \includegraphics[width=\textwidth]{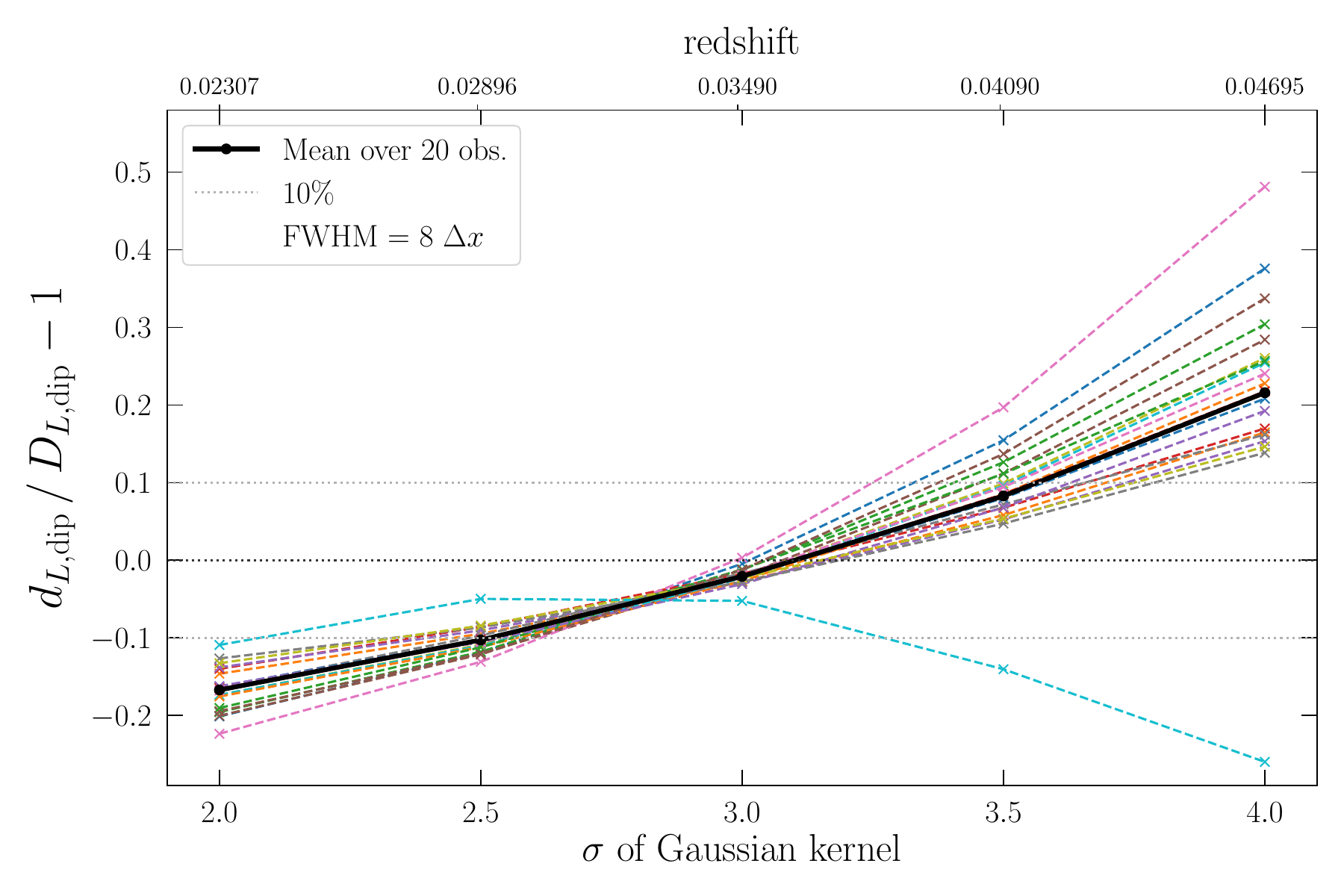}
    \end{minipage}
    \vskip\baselineskip
    \begin{minipage}[b]{0.49\textwidth}   
        \centering 
        \includegraphics[width=\textwidth]{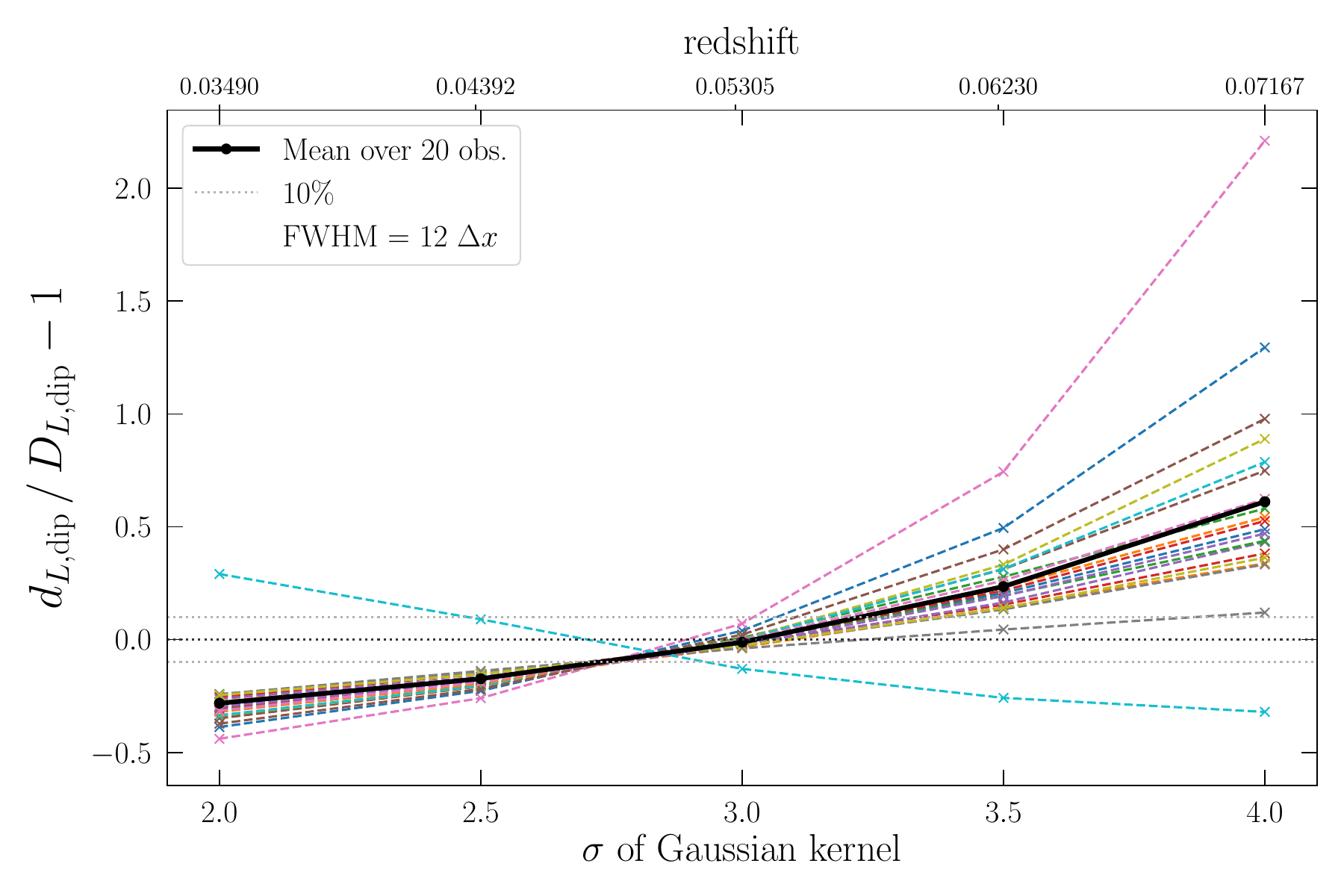}
    \end{minipage}
    \hfill
    \begin{minipage}[b]{0.49\textwidth}   
        \centering 
        \includegraphics[width=\textwidth]{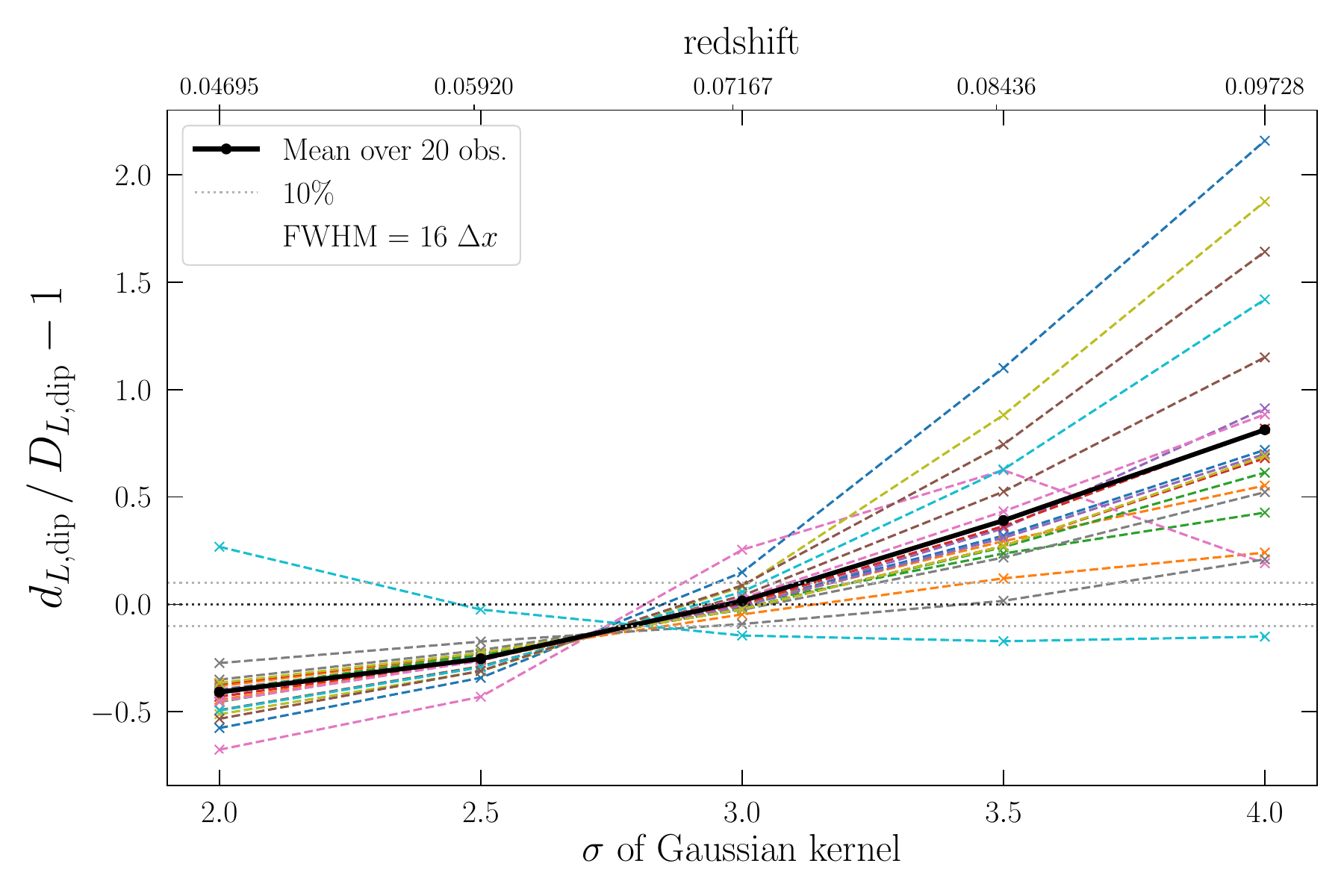}
    \end{minipage}
    \caption{The impact of changing the $n$--$\sigma$ matching with a redshift on the predicted dipole accuracy. Each panel shows a different FWHM smoothing scale of the Gaussian kernel (indicated in the legend) in the $N=256$ $L=2.56\,h^{-1}$ Gpc simulation for 20 observers (coloured curves; and their average in black). Each point in each panel represents a different matching of physical scale ($n$--$\sigma$ of the Gaussian smoothing kernel) with redshift used in the predicted and ray-traced dipole signatures.} 
    \label{fig:4panel_sigtest}
\end{figure*}

In the main text, we have chosen to match the 3--$\sigma$ width of the Gaussian kernel to an approximate redshift by comparing the predicted dipole to the ray-traced dipole. The precise procedure is the following: the quantities $\theta$ and $\rho$ are smoothed with a Gaussian kernel with full-width-half-maximum (FWHM) equal to a certain number of grid cells for that simulation. For this FWHM, we find the corresponding 3--$\sigma$ comoving distance {scale} in the simulation, which we then translate to a redshift using the FLRW relation. 
This is the redshift we use in \eqref{eq:dLdip_predic} to calculate the predicted dipole as well as the redshift {at which} we {evaluate} the ray-traced $D_L(z)$ relation, from which we {then} extract the ray-traced dipole. 
All simulations comoving distances remain very close to the FLRW relation due to the fact that the metric perturbation remains small for the entirety of the simulation, i.e. fluctuations in $\sqrt{\gamma}$ are $\mathcal{O}(10^{-5})$. For the comoving distance-redshift relation, we use a Hubble constant calculated on the final slice of the simulation as $\mathcal{H}=\frac{1}{3}\langle\theta\rangle$ where the average is taken over the entire simulation domain. 

In this Appendix we will vary the scale we translate to a redshift across 2--$\sigma$, 2.5--$\sigma$, 3--$\sigma$, 3.5--$\sigma$ and 4--$\sigma$ widths of the Gaussian kernel. For each scale, we will calculate the mismatch in the predicted versus ray traced dipole and determine which scale gives the most accurate prediction.

Figure~\ref{fig:4panel_sigtest} shows the impact of changing the physical scale, with respect to the Gaussian width, that we match with a redshift to predict the dipole signature. Each panel shows the results smoothed with a Gaussian kernel of FWHM width 4, 8, 12, and 16 (top-left to bottom-right, respectively) grid cells for the $N=256$ $L=2.56\,h^{-1}$ Gpc simulation. Each point in each panel represents replacing the 3--$\sigma$ with the $n$--$\sigma$ shown on the $x$-axis in the process described at the beginning of this section. 
The main results for this simulation, shown in Figure~\ref{fig:2p56_dLdip_amp_vs_z}, correspond to the 3--$\sigma$ points in each of the four panels. We exclude the highest-redshift point in Figure~\ref{fig:2p56_dLdip_amp_vs_z} (corresponding to $z\approx0.15$) because changing the scale makes little difference with respect to the relatively large mismatch we find at this redshift.

We conclude that overall, the choice of 3--$\sigma$ scale matching with redshift is the optimal choice. For the lower-redshift panels in Figure~\ref{fig:4panel_sigtest} (top row), the ideal choice actually is more between 3.25--3.5$\sigma$, which is responsible for the slight bias towards negative values on the $y$-axis visible in the solid curves in Figure~\ref{fig:RTvspredic_liny}. Thus, the prediction could be improved slightly by specifically tailoring the matching of smoothing scale with redshift for the particular survey redshift range of interest. While this is beyond the scope of this preliminary study, it offers hope to improve the fit in the potential application to observational data.

\section{Constraint violation}\label{appx:constraint}

The Hamiltonian constraint equation \eqref{eq:Hconstraint} is identically zero for an exact solution of Einstein's equations. For a numerical solution, such as we obtain using NR, these constraints are non-zero due to the accumulation of finite-difference error through the evolution. A small amount of constraint violation may also be introduced in the generation of initial data if some approximations to Einstein's equations are made in generating the data. Tracking the amount of constraint violation, and importantly ensuring it converges towards zero as we increase resolution, is a useful way to ensure our simulations remain trustworthy and sufficiently close to a true solution of Einstein's equations. 

We calculate the level of violation of the Hamiltonian constraint using \texttt{mescaline} and normalise it by the sum of the squares of the individual terms involved, specifically we assess $H/[H]$ where $H$ is defined in \eqref{eq:Hconstraint} and \citep[see also][]{Macpherson:2019a,Mertens:2015ttp}
\begin{equation}
    [H] \equiv \sqrt{\mathcal{R}^2 - (K_{ij}K^{ij})^2 + (K^2)^2 - (16\pi G \rho)^2}.
\end{equation}

In our simulations, the dominant source of violation arises from the evolution itself due to the high accuracy of our initial data as described in Section~\ref{sec:ICs}. While both of these sources yield fourth-order errors, the numerical error accumulates with every time step whereas the initial data is one single violation of $\mathcal{O}(\Delta x)^4$. So long as the physical situation in the simulation involved remains roughly constant, we thus expect the constraint violation to reduce as we increase numerical resolution. 

\begin{figure*}[h]
    \centering
    \includegraphics[width=0.75\textwidth]{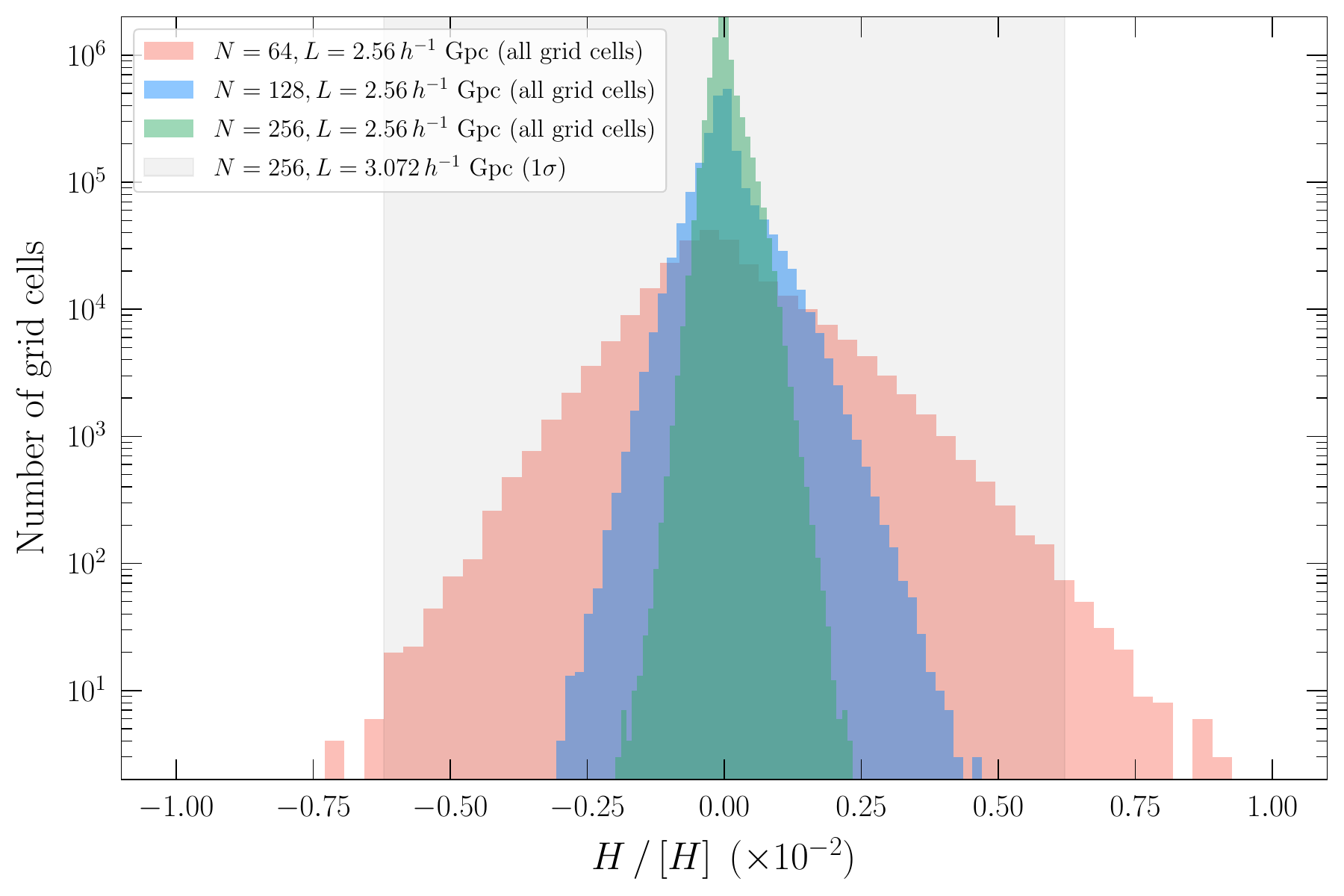}
    \caption{Hamiltonian constraint violation for the three simulations with $L=2.56\,h^{-1}$ Gpc and $N=64$ (pink), 128 (blue), and 256 (green) for all cells on each simulation final time slice. Light grey shaded region shows the $\pm 1\sigma$ spread of the violation for the $N=256$, $L=3.072\,h^{-1}$ Gpc simulation which has a higher violation due to increased levels of small-scale structures.}
    \label{fig:Hrel_viol}
\end{figure*}
Figure~\ref{fig:Hrel_viol} shows histograms of the Hamiltonian constraint violation in the three numerical test simulations with $L=2.56\,h^{-1}$ Gpc and $N=64$ (pink), 128 (blue), and 256 (green) for all grid cells in each respective simulation domain on the final time slice. This shows the level of violation on the hypersurface on which the observers are placed, and is an upper limit for the ray tracing calculations since this in general reduces with redshift (as the amplitude of structures reduces as we look back in time). For all grid cells, the level of constraint violation is below 1\% and for the highest resolution is below 0.25\%. 
The light grey shaded band in Figure~\ref{fig:Hrel_viol} shows the $\pm 1\sigma$ spread of the Hamiltonian constraint violation for the $L=3.072\,h^{-1}$ Gpc simulation with $N=256$. This simulation has significantly more small-scale structures and thus a higher level of constraint violation in general. Regardless, the violation for this simulation is at maximum $\sim 5\%$ and its $3\sigma$ width is $\sim 2\%$. 

Given both the facts that a) in a controlled situation the constraint violation reduces with resolution and b) for our `nonlinear' simulation the violation is at most a few percent, we conclude that our simulation results are reasonable solutions to Einstein's equations.

\bibliography{refs} 
\bibliographystyle{jphysicsB}

\end{document}